\documentclass[12pt]{article}
\usepackage{amssymb,amsmath}
\usepackage{epsfig}
\usepackage{epstopdf}
\usepackage{latexsym}

\def\coeff#1#2{\relax{\textstyle {#1 \over #2}}\displaystyle}

\def\IR{\mathbb{R}}
\def\ZZ{\mathbb{Z}}

\def\cO{{\cal O}}
\def\cQ{{\cal Q}}

\def\ds{\displaystyle}

\begin{document}

 \begin{titlepage}

\begin{flushright}
SPhT-T07/059
\end{flushright}

\bigskip
\bigskip
\centerline{\Large \bf Bubbles on Manifolds with a $U(1)$
Isometry}
\bigskip
\bigskip
\centerline{{\bf Iosif Bena$^1$, Nikolay Bobev$^2$ and Nicholas P.
Warner$^2$}}
\bigskip
\centerline{$^1$ Service de Physique Th\'eorique, }
\centerline{CEA Saclay, 91191 Gif sur Yvette, France}
\bigskip
\centerline{$^2$ Department of Physics and Astronomy}
\centerline{University of Southern California} \centerline{Los
Angeles, CA 90089, USA}
\bigskip
\centerline{{\rm iosif.bena@cea.fr, bobev@usc.edu,
warner@usc.edu} }
\bigskip \bigskip

\begin{abstract}

  We investigate the construction of five-dimensional, three-charge supergravity
  solutions that only have a rotational $U(1)$ isometry. We show that such
  solutions can be obtained as warped compactifications with a
  singular ambi-polar hyper-K\"ahler base space and singular warp
  factors.  We show that the complete solution is regular around the
  critical surface of the ambi-polar base.  We illustrate this by
  presenting the explicit form of the most general supersymmetric
  solutions that can be obtained from an Atiyah-Hitchin base space and
  its ambi-polar generalizations.  We make a parallel analysis using an
ambi-polar generalization of the Eguchi-Hanson base space
  metric.  We also show how the bubbling procedure applied to the
  ambi-polar Eguchi-Hanson metric can convert it to a global $AdS_2
  \times S^3$ compactification.

\end{abstract}

\end{titlepage}

\section{Introduction}

Geometric transitions have proven to be an essential part of
understanding string theory and strongly coupled quantum field
theories.  It has also become evident that such transitions will
play a central role in understanding the geometry of microstates
of the three-charge black hole or black ring in five dimensions.
Indeed, one can argue \cite{Bena:2005va} that a very large number
of horizonless three-charge brane configurations, when brought to
strong effective coupling, undergo a geometric transition and
become smooth horizonless geometries with black-hole or black-ring
charges. The black hole and black ring charges come entirely from
fluxes wrapping topologically non-trivial cycles, or bubbles. All
these ``bubbling black hole'' geometries are dual to states of the
conformal field theory describing the three-charge black hole, and
their physics strongly supports the idea that black holes in
string theory are not fundamental objects, but rather effective
thermodynamic descriptions of a huge number of horizonless
configurations (see \cite{samir,Bena:2007kg} for a review).

The simplest starting point for the construction of three-charge
geometries is M-theory, where one takes the eleven-dimensional
supersymmetric metrics to have the form \cite{Bena:2004de,harvey}:
\begin{eqnarray}
 ds_{11}^2  = ds_5^2 & + &    \left(Z_2 Z_3  Z_1^{-2}  \right)^{1\over 3}
 (dx_5^2+dx_6^2) \nonumber \\
 & + & \left( Z_1 Z_3  Z_2^{-2} \right)^{1\over 3} (dx_7^2+dx_8^2)    +
  \left(Z_1 Z_2  Z_3^{-2} \right)^{1\over 3} (dx_9^2+dx_{10}^2) \,.
\label{elevenmetric}
\end{eqnarray}
The six coordinates, $x_A$, parameterize the compactification
torus, $T^6$, and the five-dimensional space-time metric has the
form:
\begin{equation}
ds_5^2 ~\equiv~ - \left( Z_1 Z_2  Z_3 \right)^{-{2\over 3}}
(dt+k)^2 + \left( Z_1 Z_2 Z_3\right)^{1\over 3} \, ds_4^2 \,.
\label{fivemetric}
\end{equation}
When constructing black-hole or  black-ring solutions
\cite{blackring1,blackring2,Bena:2004de,blackring3,jerome}, the
spatial ``base metric,'' $ds_4^2$, is usually taken to be that of
flat $\IR^4$,  however supersymmetry is preserved if one has any
hyper-K\"ahler metric  on the the base \cite{5dsols}.

To construct bubbling solutions corresponding to five-dimensional
black holes and black rings, the asymptotic structure of $ds_4^2$
must still be that of $\IR^4$ and the conventional wisdom
suggested that this implies that the hyper-K\"ahler base metric
must be flat $\IR^4$, globally. The breakthrough came via
\cite{Giusto:2004kj,Bena:2005va,Berglund:2005vb}, where it was
realized that the base metric could be ``ambi-polar,'' that is, it
could change its overall signature from $(+,+,+,+)$ to $(-,-,-,-)$
in some regions. The warp factors, $Z_I$, are also singular and
change sign, but the complete five-dimensional (or
eleven-dimensional) solution is still a physical, Lorentzian
metric. This opens up a vast number of new possibilities for
constructing bubbling black holes using four-dimensional
hyper-K\"ahler base metrics.

The construction of the most general solutions proceeds in several
steps using the linear system of ``BPS equations''
\cite{Bena:2004de}. One first chooses the ambi-polar,
hyper-K\"ahler base metric.  This base will generically have
non-trivial two-cycles (``bubbles'') with moduli that determine
the size and orientation of the cycles.  Dual to these bubbles are
normalizable\footnote{Here we are, to some extent, misusing the
term ``normalizable:''  The basic ``component'' fluxes are
normalizable in that they fall off sufficiently rapidly at
infinity but the component fluxes are divergent on the critical
surfaces where the metric of the base changes sign. However, the
physical fluxes get two contributions in which these divergences
cancel and so the ultimate flux that we construct will be
normalizable.}, self-dual, harmonic
 two-forms (the compact cohomology).  There are also three non-normalizable,
anti-self-dual, harmonic two-forms and these correspond to the
three complex structures of the hyper-K\"ahler base\footnote{If
one reverses
  the orientation of the base then the self-dual and anti-self-dual
  forms will, of course, be interchanged.}.  The normalizable harmonic
two-forms determine the electromagnetic fluxes, which in turn
source the warp factors, $Z_I$.  Finally, the warp factors and the
fluxes combine to give the source in the linear equation for the
angular momentum vector, $k$, in (\ref{fivemetric}).

This procedure is easiest implemented when one considers
ambi-polar Gibbons-Hawking (GH) metrics, which are hyper-K\"ahler
metrics with a tri-holomorphic $U(1)$ symmetry\footnote{
Tri-holomorphic means that
  the $U(1)$ preserves all three complex structures of the
  hyper-K\"ahler metric.}. All five-dimensional, supersymmetric
solutions with Gibbons-Hawking base can be written in terms of
several harmonic functions \cite{5dsols,jerome,BRtaub}; choosing
an ambi-polar GH metric (with positive and negative GH charges)
and specific harmonic functions ensures that the resulting
``bubbling'' solutions are horizonless, smooth, and have black
hole and black ring charges \cite{Bena:2005va, Berglund:2005vb,
Saxena:2005uk, Bena:2006is,
  Bena:2006kb, Cheng:2006yq, Ford:2006yb}.

At the last step of the construction, some of the moduli of the
bubbles have to be fixed so as to remove closed time-like curves
(CTC's). This step has a simple, physical interpretation: The
ambi-polar base metric typically arises because one has used both
positive and negative geometric charges ({\it e.g.}  GH charges)
that tend to attract one another and would cause the space to
collapse if they were not stabilized in some manner.  If one seeks
BPS solutions without including any physical stabilizing
mechanism, the instability will manifest itself through the
appearance of closed time-like curves (CTC's).   On the other
hand, a flux threading a cycle tends to cause that cycle to
expand.  Hence, by distributing fluxes on the space one can
balance the attraction of the geometric charges against the
expansion caused by fluxes.  The bubbles thus settle down at a
size where the attractive and expansion forces cancel and the
result is a stable BPS solution free of CTC's.  One generically
finds that the sizes of the bubbles are set in terms of the fluxes
threading them.  Other bubble moduli, like orientations, remain
free.  The equations that express this balance of fluxes and
bubble sizes are called the ``bubble equations''
\cite{Bena:2005va,Berglund:2005vb}.

Studying solutions that have an ambi-polar Gibbons-Hawking base
has quite a few advantages: the construction of these solutions is
straightforward, the solutions can be related to four-dimensional,
multi-centered ``D6 - $\overline{\rm D6}$'' solutions
\cite{denef,BRtaub,4d5d} \footnote{In such compactifications, the
bubble equations correspond  to the
  ``integrability equations'' discussed in \cite{denef}.}, they can
arise from the geometric transition of circular three-charge
supertubes \cite{3chsuper,Bena:2005va}, and they can describe
microstates both of zero-entropy black holes and black rings
\cite{Bena:2006is}, or of black holes and black rings with
macroscopic horizons \cite{Bena:2006kb,deepring}.

Despite the remarkable results that have been obtained using
Gibbons-Hawking geometries, such metrics represent a major
restriction. In particular, they all have a translational
(tri-holomorphic) $U(1)$ isometry \cite{Gibbons:1987sp}, which is
a combination of the two $U(1)$'s in the $\IR^2$ planes that make
up the $\IR^4$ in the asymptotic region\footnote{Alternatively,
one can see
  that the tri-holomorphic $U(1)$ necessarily lies in one of the
  $SU(2)$ factors of $SO(4) \equiv (SU(2) \times SU(2))/\ZZ_2$.}.
Thus, bubbling solutions with a GH base cannot capture quite a
host of interesting physical processes that do not respect this
symmetry, like the merger of two BMPV black holes, or the
geometric transition of a three-charge supertube of arbitrary
shape.  In \cite{Bena:2005va} it was argued that this geometric
transition results in bubbling solutions that have an ambi-polar
hyper-K\"ahler base, and that depend on a very large number of
arbitrary continuous functions.  Counting these solutions is a way
to prove the general conjecture that black holes are ensembles of
smooth horizonless configurations. It is therefore of great
interest to construct and understand them.

In this paper we will take a step in this direction by considering
such metrics that also have a general $U(1)$ isometry. These
metrics are much less restrictive than GH-based metrics. Moreover,
they could also arise from the geometric transition of supertubes
that preserve a rotational $U(1)$, and hence could also depend on
an arbitrarily large number of continuous functions. Constructing
and counting such solutions is also of interest in the program to
prove that black holes are ensembles of smooth horizonless
configurations.  Even if the entropy in these symmetric
configurations will be smaller than the entropy of the black hole,
it might give some insight into the structure and charge
dependence of the most general, non-symmetric configuration.

An important feature of all bubbled solutions is that the
ambi-polar base space and  the fluxes dual to the homology are
singular on the critical surfaces where the metric changes sign.
For ambi-polar GH spaces it was possible to use the explicit
solutions to show that all these singularities were canceled and
the final result was a regular, five-dimensional space-time
background in M-theory. Our analysis here will illustrate how this
happens for the general $U(1)$-invariant bubbled background, and
this work suggests that  the most general bubbling geometries will
have also this property.

Before beginning, we would like to stress that constructing
solutions that only have a rotational $U(1)$ is a rather tedious
and challenging task. For classical black holes and black rings,
only two such solutions exist: one describing a black ring with an
arbitrary charge density \cite{ChihWei1}, and one describing a
black ring with a black hole away from the center of the ring
\cite{ChihWei2}. In this paper we will succeed in constructing the
first explicit {\it bubbling} solution in this class, using a base
that is a generalization of the Atiyah-Hitchin metric\footnote{Our
solutions might also be useful to construct black holes or black
rings in the Atiyah-Hitchin space, although this has not been the
focus of this paper.}. Nevertheless, the most general bubbling
solutions that only have a rotational $U(1)$ invariance will be
much more complicated, and perhaps even impossible to write down
explicitly.

\section{Prelude}

It has been known for a long time that hyper-K\"ahler metrics with
a generic (rotational)  $U(1)$ isometry can be obtained by solving
the $SU(\infty)$ Toda equation
\cite{Boyer:1982mm,DasGegenberg,Bakas:1996gf}.  The coordinates
can be chosen so that the metric takes the form:
\begin{equation}
ds^2_4 ~=~ V^{-1}\, (d\tau+A_idx^i)^2 ~+~ V\gamma_{ij}\,dx^idx^j
\,, \label{Todametric}
\end{equation}
with $\gamma_{ij} = 0$ for $i \ne j$ and
\begin{equation}
\gamma_{11}~=~\gamma_{22}~=~e^\nu \,, \qquad \gamma_{33}~=~1 \,,
\label{threemetric}
\end{equation}
for some function, $\nu$.  The function, $V$, and the vector
field, $A$, are given by
\begin{equation}
V~=~ \partial_z \nu \,,  \qquad A_1 ~=~ \partial_y\nu \,,  \qquad
A_2~=~ -\partial_x \nu\,,  \qquad A_3~=~ 0 \,, \label{VAform}
\end{equation}
and the function $\nu$ must satisfy:
\begin{equation}
\partial_x^2\,\nu~+~  \partial_y^2\,\nu  ~+~ \partial_z^2\,(e^{\nu})~=~0 \,.
\label{TodaEqn}
\end{equation}
This equation is called the $SU(\infty)$ Toda equation, and may be
viewed as a continuum limit of the $SU(N)$  Toda equation.  Even
though the $SU(\infty)$ Toda equation is integrable, surprisingly
little is know about its solutions, and there appears to be no
known analog of the known soliton solutions of  the $SU(N)$ Toda
equation. On the other hand, the metric is determined in terms of
a single function and (\ref{Todametric}) is a relatively mild
generalization of the Gibbons-Hawking metrics.

Our purpose here is to construct three-charge solutions based upon
ambi-polar hyper-K\"ahler metrics with generic
(non-tri-holomorphic) $U(1)$ isometries. We will do this in two
different ways, first by building such solutions using a general
metric of the form (\ref{Todametric}) on the base space.   We will
then consider the Atiyah-Hitchin metric:  This metric has an
$SO(3)$ isometry, but none of the $U(1)$ subgroups is
tri-holomorphic.  Just as with the GH metric, the metric
(\ref{Todametric}), is ambi-polar if we allow $V = \partial_z \nu$
to change sign.  Thus the primary issue of regularity in the
five-dimensional metric arises on the {\it critical surfaces}
where $V=0$.  While we will not be able to construct general
solutions as explicitly as can be done for GH metrics, we will
show that the five-dimensional metric is regular and Lorentzian in
the neighborhood of these critical surfaces.

The standard Atiyah-Hitchin metric \cite{Atiyah:1985dv} arises as
the solution of a first order, non-linear Darboux-Halphen system
for the three metric coefficient functions.  This system is
analytically solvable in terms of the solution of a single, second
order linear differential equation.  Indeed, the solutions of the
latter equation are expressible in terms of elliptic functions.
The standard practice is to choose the solution of this linear
equation so that the metric functions are regular, and the result
is a smooth geometry that closes off at a non-trivial ``bolt,'' or
two-cycle in the center.  We will show that if one selects the
most general solution of the linear differential equation, then
one obtains an ambi-polar generalization of the Atiyah-Hitchin
metric.  Moreover, one can set up regular, cohomological fluxes on
the two-cycle and the resulting warp factors render the
five-dimensional metric perfectly smooth and regular across the
critical surface.

The ambi-polar Atiyah-Hitchin metric actually continues through
the bolt and initially appears to have two regions, one on each
side of the bolt, that are asymptotic to $\IR^3 \times S^1$.  It
thus looks like a wormhole.  Unfortunately, the solution cannot be
made regular on {\it both} its asymptotic regions.  Indeed, upon
imposing asymtotic flatness on one side of the wormhole, one finds
that the warp factors change sign twice, once on the critical
surface and again as one enters one of the asymptotic regions.
Thus the critical surface is regular, but there is another
potentially singular region elsewhere. However, we find that if we
tune the flux through the bubble to exactly the proper value, one
can pinch off the metric just as the warp factors change sign for
the second time.  The result is a Lorentzian metric, that extends
smoothly through the critical surface ($V=0$).  The pinching off
of the metric does however result in a curvature singularity that
is very similar to the one encountered in the Klebanov-Tseytlin
solution \cite{Klebanov:2000nc}. We will argue that the
singularity of our new, non-trivial BPS solution is also a
consequence of the very high level of symmetry, and that it will
be resolved via a mechanism similar to that in
\cite{Klebanov:2000hb}.

We also consider solutions based upon an ambi-polar generalization
of the Eguchi-Hanson metric, obtained by making an analytic
continuation of the standard Eguchi-Hanson metric, and extending
the range of one of the coordinates\footnote{The unextended
version of this metric was
  also discussed in the original Eguchi and Hanson paper
  \cite{Eguchi:1978xp}, but was discarded because it is singular.}.
The singular structure of this metric is precisely what is needed
to render it ambi-polar. Hence, upon adding fluxes and warp
factors this metric gives us regular five-dimensional solutions
that have similar features to the bubling Atiyah-Hitchin solution.
There is also one surprise: One of the Eguchi-Hanson ``wormhole''
solutions is completely regular everywhere and is nothing other
than the global $AdS_2 \times S^3$ Robinson-Bertotti solution.

In Section 3 we will review the BPS equations and discuss their
solution for a metric with a generic $U(1)$ isometry.  In Section
4 we will review the properties and structure of the
Atiyah-Hitchin metric. Section 5 is devoted to the explicit
solution of the system of BPS equations for the Atiyah-Hitchin
metric and its ambi-polar generalization, and a discussion on
regularity and CTC's.  In Section 6 we present the bubbled
solutions on the generalized (ambi-polar) Eguchi-Hanson background
and we show how they closely parallel the results for the
generalized (ambi-polar) Atiyah-Hitchin solution. We also show how
to obtain {\it global} $AdS_2 \times S^3$ as a bubbling
Eguchi-Hanson solution. Finally, Section 7 contains our
conclusions and a discussion of possible future work.

\section{The general $U(1)$ invariant  geometries}
\label{GenGeoms}

\subsection{The BPS equations}

The supersymmetric, BPS solutions to M-theory with metric given by
(\ref{elevenmetric}) have Maxwell three-form potential given by:
\begin{equation}
C^{(3)}  = A^{(1)} \wedge dx_5 \wedge dx_6 ~+~  A^{(2)}   \wedge
dx_7 \wedge dx_8 ~+~ A^{(3)}  \wedge dx_9 \wedge dx_{10}  \,,
\label{Cfield:ring}
\end{equation}
where the $A^{(I)}$, $I=1,2, 3$, are one-form Maxwell potentials
in the five-dimensional space-time and depend only upon the
coordinates, $y^\mu$, $\mu =1,\dots, 4$, that parameterize the
spatial directions of $ds_4$.  It is convenient to introduce the
Maxwell
 ``dipole field strengths,''    $\Theta^{(I)}$,  obtained by removing the contributions
 of the electrostatic potentials:
\begin{equation}
\Theta^{(I)} ~\equiv~  d A^{(I)} + d\big(  Z_I^{-1} \, (dt +k)
\big)  \,. \label{Thetadefn}
\end{equation}

The most general supersymmetric configuration is then obtained by
solving the {\it BPS equations}:
\begin{eqnarray}
 \Theta^{(I)}  &~=~&  \star_4 \, \Theta^{(I)} \label{BPSeqn:1} \,, \\
 \nabla^2  Z_I &~=~&  {1 \over 2  }  C_{IJK} \star_4 (\Theta^{(J)} \wedge
\Theta^{(K)})  \label{BPSeqn:2} \,, \\
 dk ~+~  \star_4 dk &~=~&  Z_I \,  \Theta^{(I)}\,,
\label{BPSeqn:3}
\end{eqnarray}
where $\star_4$ is the Hodge dual taken with respect to the
four-dimensional base metric, $ds_4^2$, and for the structure
constants are given by $C_{IJK} ~\equiv~ |\epsilon_{IJK}|$. The
{\it BPS equations} generalize trivially to more general $U(1)^N$
five-dimensional ungauged supergravities.

The first step in solving this linear system is to identify the
self-dual, harmonic two-forms, $\Theta^{(I)}$.  In a K\"ahler
manifold this is, at least theoretically, straightforward because
such two-forms are related to the moduli of the metric.   For a
hyper-K\"ahler metric there are three complex structures,
$J_{(i)}$, $i=1,2,3$, and given a harmonic two-form, $\omega$, one
can define three symmetric tensors via:
\begin{equation}
h^{(i)}_{\mu \nu}~\equiv~    {J_{(i)\,\mu}}^\rho \,
\omega_{\rho\nu}  ~-~
 {J_{(i)\,\nu}}^\rho \, \omega_{\mu \rho}\,.
\label{Fluctuations}
\end{equation}
These tensors may be viewed as metric perturbations and as such
they represent perturbations that preserve the hyper-K\"ahler
structure.  In particular, they are zero modes of the Lichnerowicz
operator.

For the metric (\ref{Todametric}) it is convenient to introduce
vierbeins:
\begin{equation}
 \hat{e}^1=V^{-1/2}(d\tau+ A_idx^i) \,, \qquad
\hat{e}^2=V^{1/2}e^{\nu/2}dx \,, \qquad
\hat{e}^3=V^{1/2}e^{\nu/2}dy \,, \qquad \hat{e}^4=V^{1/2}dz  \,,
\end{equation}
and introduce a basis for the self-dual and anti-self dual two
forms:
\begin{eqnarray}
 \Omega_{\pm}^{(1) }&=& (d\tau+A_2dy)\wedge dx ~\pm~
Vdy\wedge dz  ~=~  e^{-\nu/2}(\hat e^1\wedge\hat e^2~\pm~ \hat e^3\wedge\hat e^4)\,, \\
\Omega_{\pm}^{(2)} &=& (d\tau+A_1dx)\wedge dy ~\pm~
V dz\wedge dx ~=~  e^{-\nu/2}(\hat e^1\wedge\hat e^3~\pm~\hat e^4\wedge\hat e^2) \,, \\
\Omega_{\pm}^{(3)} &=& (d\tau+A_1dx+A_2dy)\wedge dz ~\pm~
e^{\nu}Vdx\wedge dy~=~ (\hat e^1\wedge\hat e^4~\pm~\hat
e^2\wedge\hat e^3)\,.
\end{eqnarray}
The three K\"{a}hler forms are then given by \cite{Bakas:1996gf}:
\begin{eqnarray}
J_{(1)} &=& e^{\nu/2}\cos\Big(\frac{\tau}{2}\Big)\,
\Omega_{-}^{(1)} ~+~ e^{\nu/2}\sin\Big(\frac{\tau}{2}\Big)\, \Omega_{-}^{(2)\,, }\\
J_{(2)} &=& e^{\nu/2}\sin\Big(\frac{\tau}{2}\Big)\,
\Omega_{-}^{(1)} ~-~ e^{\nu/2} \cos\Big(\frac{\tau}{2}\Big )\Omega_{-}^{(2)}\,,\\
J_{(3)} &=& \Omega_{-}^{(3)} \,,
\end{eqnarray}
and they satisfy the proper quaternionic algebra:
\begin{equation}
 {J_{(i)\,\mu}}^\rho \,  {J_{(j)\,\rho}}^\nu ~=~
\delta_{ij}\, \delta_\mu^\nu ~-~  \varepsilon_{ijk}\,
{J_{(k)\,\mu}}^\nu \,.
\end{equation}

Following \cite{Bena:2005va}, we make an Ansatz for the harmonic,
self-dual field strengths, $\Theta^{(I)}$:
\begin{equation}
\Theta^{(I)}~=~  \sum_{a=1}^{3}\partial_{a}(\dot{\nu}^{-1}K^{I})\,
\Omega_{+}^{(a)} \,,
 \label{Thetaform}
\end{equation}
where the dot represents derivative with respect to $z$.   We then
find that the $K^I$ must satisfy the linearized Toda equation (it
follows from (\ref{TodaEqn}) that $\dot{\nu}$ also solves this
equation):
\begin{equation}
{\cal L_T} \, K^{I} ~\equiv~ \partial_x^2K^{I}~+~
\partial_y^2K^{I} ~+~
\partial_z^2(e^{\nu}K^{I})~=~ 0  \,.
 \label{linToda}
\end{equation}
For later convenience, we note  that there are relatively simple
vector potentials such that $\Theta^{(I)} = dB^{(I)}$:
\begin{equation}
B^{(I)} ~\equiv~  \dot \nu^{-1}\,  K^{I}  \, (d\tau + A) ~+~
{\vec{\xi}}^{\,(I)} \cdot d \vec x \,, \label{Bpot}
\end{equation}
where
\begin{equation}
(\vec  \nabla \times {\vec{\xi}}^{\,(I)})^j  ~=~ - \partial_i
\big(\gamma^{ij} e^\nu K^{I}\big) \,. \label{xidefn}
\end{equation}
Hence, ${\vec{\xi}}^{\,(I)}$ is a vector potential for magnetic
monopoles located at the singular points of $K^{I} $.

Since the $K^I$ satisfy the linearized Toda equation, we see the
direct relationship between the harmonic forms and linearized
fluctuations of the metric. In practice, (\ref{Fluctuations}) and
(\ref{Thetaform}) do not yield exactly the same result as the
direct substitution of fluctuations in $\nu$ into
(\ref{Todametric})  but they are equivalent up to infinitesimal
diffeomorphisms. For example, the metric fluctuation obtained from
using (\ref{Thetaform}) and $J_{(3)}$ in (\ref{Fluctuations}) is
identical with the metric fluctuation, $\nu \to \nu + \epsilon
K^I$, combined with the infinitesimal diffeomorphism, $ z \to z -
\epsilon \dot \nu^{-1} K^I$.

The second BPS equation reduces to:
\begin{equation}
 {\cal L}  Z_I ~=~  \dot \nu \, e^{\nu}\,  C_{IJK}\, \gamma^{ij}  \partial_i
\Big( {K^J \over \dot  \nu} \Big)\, \partial_j  \Big( {K^K \over
\dot  \nu}\Big)\,, \label{SecBPS}
\end{equation}
where $\gamma_{ij}$ is the  three-metric in (\ref{threemetric})
and ${\cal L}$ is given by:
\begin{equation}
 {\cal L}  \, F  ~\equiv~ \dot \nu \, e^{\nu}\,  \nabla_\gamma^2 F  ~=~
 \partial_x^2 F+~ \partial_y^2 F ~+~  \partial_z (e^{\nu}\partial_z  F)  \,.
 \label{Laplacian}
\end{equation}
The operator, $\nabla_\gamma^2$, denotes the Laplacian in the
metric $\gamma_{ij}$.

The natural guess for the solution is to follow, once again,
\cite{Bena:2005va} and try:
\begin{equation}
Z_I ~\equiv ~ \coeff{1}{2}  \, C_{IJK} \, \dot \nu^{-1}\,K^J K^K
~+~ Z_I^{(0)} \,.
 \label{Zzerodefn}
\end{equation}
One then finds that $Z_I^{(0)}$ is not a solution of the
homogeneous equation, but
\begin{equation}
 {\cal L} \, Z_I^{(0)}  ~=~  -\partial_z \big( \coeff{1}{2}\, e^{\nu}  \, C_{IJK} \,  K^J K^K \big) \,.
\label{Zzeroeqn}
\end{equation}
Intriguingly, one can also check that:
\begin{equation}
 {\cal L}_T \big( \coeff{1}{2}  \, C_{IJK} \, \dot \nu^{-1}\,K^J K^K \big) ~=~
 \dot \nu \, e^{\nu}\,  C_{IJK}\, \gamma^{ij}  \partial_i
\Big( {K^J \over \dot  \nu} \Big)\, \partial_j  \Big( {K^K \over
\dot  \nu}\Big)\,, \label{LinTodaYI}
\end{equation}
where $ {\cal L}_T $ is the linearized Toda operator
(\ref{linToda}) and so one has the explicit solution but to the
wrong equation.

The important point, however, is that the source on the right-hand
side of
 (\ref{Zzeroeqn}) is regular as  $\dot \nu \to 0$, and so $Z_I^{(0)}$ is regular
 on any critical surface where one has $\dot \nu=0$.

To solve the last BPS equation for the angular momentum vector,
$k$, we make the Ansatz:
\begin{equation}
k ~=~ \mu\, ( d\tau + A   ) ~+~ \omega \,, \label{kansatz}
\end{equation}
where $\omega$ is a one form in the three-dimensional space
defined by $(x,y,z)$.  Define yet another linear operator:
\begin{equation}
\widetilde {\cal L}  \, F  ~\equiv~  e^{\nu} \gamma^{ij} \,
\partial_i \, \partial_j F~=~
\partial_x^2 F+~ \partial_y^2 F ~+~  e^{\nu}\, \partial_z^2 F \,,
 \label{NewLinOp}
\end{equation}
and then one finds that $\mu$ and $\omega$ must satisfy:
\begin{equation}
\widetilde {\cal L}  \, \mu  ~=~   \dot \nu^{-1}\, \partial_i
\Big( \dot \nu\, e^{\nu}\,
 \gamma^{ij} \, \sum_{I=1}^3 \, Z_I\,\partial_j\Big( {K^I \over \dot \nu} \Big)\Big) \,,
 \label{mueqn}
\end{equation}
and
\begin{equation}
(\vec\nabla \times \vec\omega)^i ~=~ \dot\nu \,
e^{\nu}\gamma^{ij}\partial_j\mu ~-~ \mu \,\partial_j(e^{\nu}
\gamma ^{ij}\dot\nu) ~-~  \dot\nu \, e^{\nu} \, Z_I\, \gamma
^{ij}\,
\partial_j \Big( {K^I \over \dot\nu}\Big)  \,.
 \label{omegaeqn}
\end{equation}
Note that the integrability of the equation for $\omega$ is
precisely the equation (\ref{mueqn}) for $\mu$, provided that one
also uses the fact that $\nu$ satisfies (\ref{TodaEqn}).  The
structure of these equations also closely parallels those
encountered for a GH base metric
\cite{Bena:2005va,Berglund:2005vb}.

Once again one can try a form of the solution based upon the
results for GH spaces. Define $\mu_0$ by:
\begin{eqnarray}
\mu &=&\coeff {1}{2} \, \dot \nu^{-1} \, Z_I \, K^I ~-~
\coeff{1}{12}\,  \dot \nu^{-2} \,
C_{IJK}\,  K^I K^J K^K   ~+~  \mu_0 \nonumber \\
 &=&\coeff {1}{2} \, \dot \nu^{-1} \, Z_I^{(0)}  \, K^I ~+~ \coeff{1}{6}\,  \dot \nu^{-2} \,
C_{IJK}\,  K^I K^J K^K   ~+~  \mu_0\,, \label{mures}
\end{eqnarray}
and one then finds that $\mu_0$ must satisfy:
\begin{equation}
\widetilde {\cal L}  \, \mu_0  ~=~ - \coeff{1}{2}\, e^\nu \, K^I
\, \partial_z Z_I^{(0)}  ~+~ \coeff{1}{12}\, e^\nu \,   C_{IJK}\,
K^I K^J K^K\,.
 \label{muzeroeqn}
\end{equation}
Again note that the source is regular as $\dot \nu \to 0$ and so
$\mu_0$ will be similarly regular as  $\dot \nu \to 0$.

Finally, if one substitutes these expressions for $Z_I$ and $\mu$
into (\ref{omegaeqn}), one obtains:
\begin{eqnarray}
(\vec\nabla \times \vec\omega)^i &=& \dot\nu \,
e^{\nu}\gamma^{ij}\partial_j\mu_0 ~-~ \mu_0 \,\partial_j(e^{\nu}
\gamma ^{ij}\dot\nu) ~+~  \coeff{1}{2}\, K^I \partial_j
\big(e^{\nu}\gamma^{ij}Z_I^{(0)} \big) ~-~
\coeff{1}{2}\,e^{\nu}\gamma^{ij}Z_I^{(0)}
\partial_j K^I  \nonumber \\
&& ~-~  \coeff{1}{6}\, \delta^i_3  \, e^\nu \,   C_{IJK}\,  K^I
K^J K^K \,,
 \label{niceomegaeqn}
\end{eqnarray}
where the $\delta^i_3$ means that the last term only appears for
$i=3$. Note that $\vec \omega$ has sources that are regular as
$\dot \nu \to 0$ and so $\vec \omega$ will be regular on critical
surfaces.

Therefore, in this more general class of metrics, we cannot find
the solutions to the BPS equations as explicitly as one can for GH
base metrics.  However, one can completely and explicitly
characterize the singular parts of the solutions as one approaches
critical surfaces where $\dot \nu \to 0$.

\subsection{Regularity on the critical surfaces}

Consider the behavior of the metric (\ref{fivemetric}) as $\dot
\nu \to 0$. The warp factors, $Z_I$ diverge as $\dot \nu^{-1}$,
$\mu$ diverges as $\dot \nu^{-2}$ and so the only potentially
divergent part of the metric is:
\begin{equation}
- \big( Z_1 Z_2  Z_3 \big)^{-{2\over 3}}  \mu^2 (d \tau + A)^2 +
\big( Z_1 Z_2  Z_3 \big)^{1\over 3} \, \dot \nu^{-1}\, (d \tau +
A)^2 ~=~  \big( Z_1 Z_2  Z_3\, \dot  \nu^3 \big)^{-{2\over 3}} \,
{\cal Q} \,(d \tau + A)^2  \,,
 \label{dangerpart}
\end{equation}
where
\begin{equation}
  {\cal Q} ~\equiv~ Z_1 \,Z_2 \, Z_3\, \dot \nu ~-~ \mu^2\, \dot \nu^2 \,.
 \label{Qdefn}
\end{equation}
Every other part of the metric has a finite limit as $\dot \nu \to
0$.  Since $( Z_1 Z_2  Z_3\, \dot  \nu^3)$ is finite as $\dot \nu
\to 0$, we need to show that ${\cal Q}$ is finite.  Using
(\ref{Zzerodefn}) and (\ref{mures}) one has
\begin{eqnarray}
  {\cal Q} & = &\dot \nu^{-2} \, \Big[ \big(K^2 K^3  ~+~ \dot \nu\, Z_1^{(0)}\big)
  \big(K^1 K^3  ~+~ \dot \nu\, Z_2^{(0)}\big)\big(K^1 K^2  ~+~
  \dot \nu\, Z_3^{(0)}\big) \nonumber\\
  && \qquad \qquad \qquad~-~ \big(K^1 K^2 K^3  ~+~
  \coeff{1}{2} \, \dot \nu \, Z_I^{(0)} K^I ~+~  \dot \nu^2 \mu_0\big)^2   \Big] \nonumber \\
  &\to & \big(Z_1^{(0)}  Z_2^{(0)}\, K^1  K^2 + Z_1^{(0)}  Z_3^{(0)}\, K^1  K^3 +
  Z_2^{(0)}  Z_3^{(0)}\, K^2  K^3\big) \nonumber \\
  && \qquad \qquad \qquad~-~ \coeff{1}{4} \, \big(\dot \nu \, Z_I^{(0)} K^I \big)^2~-~
  2\, (K^1 K^2 K^3)\,\mu_0 \,,
 \label{Qdivpart}
\end{eqnarray}
as $\dot \nu \to 0$.  Thus the metric is finite on the critical
surfaces.  To avoid CTC's, ${\cal Q}$ must also be positive
everywhere and, as with solutions on GH base metrics, this will
depend upon the details of particular solutions.

The Maxwell fields are also regular on the critical surfaces.
From (\ref{Thetaform}) we see that the $\Theta^{(I)}$ are, in
fact, singular on the critical surfaces, however from
(\ref{Thetadefn}) and (\ref{Bpot}) we see that the complete
Maxwell fields are given by:
\begin{equation}
A^{(I)} ~=~   - Z_I^{-1} \,\big(dt + \mu\, ( d\tau + A  ) + \omega
\big)   ~+~ \dot \nu^{-1}\,  K^{I}  \, (d\tau + A) ~+~
{\vec{\xi}}^{\,(I)} \cdot d \vec x \,. \label{Apots}
\end{equation}
As we remarked earlier, $\omega$ is regular on the critical
surfaces and the vectors, ${\vec{\xi}}^{\,(I)}$, defined by
(\ref{xidefn}) are similarly regular. The only possible singular
terms are thus
\begin{eqnarray}
A^{(I)} &~\sim~&  \big(\dot \nu^{-1}\, K^I  ~-~  Z_I^{-1} \mu
\big) \,
(d \tau +A) \nonumber \\
& ~\sim~& \dot \nu^{-1}\, \big( K^I ~-~ \big( \coeff{1}{2}\,
C_{IJK}\,   K^J \,K^K\big)^{-1}\, K^1 \,  K^2 \,K^3 \big) \, (d
\tau +A)~=~0\,. \label{AIform}
\end{eqnarray}
Thus the $A^{(I)}$ are regular on the critical surfaces.

\subsection{Asymptotia}

We would like the four dimensional base metric to be asymptotic to
$\IR^4$ and there are several ways to arrange this, depending upon
how the $U(1)$ defined by $\tau$-translations acts in $\IR^4$.
The simplest is to take $\nu \sim \log(z)$ and then:
\begin{equation}
ds^2_4 ~\sim~ z\,  d\tau^2 ~+~ z^{-1} dz^2 ~+~ dx^2 ~+~ dy^2
~=~ dr^2 ~+~  r^2\,  d\phi^2   ~+~ dx^2 ~+~ dy^2 \,,
\label{baseasympa}
\end{equation}
where $z = \frac{1}{4} r^2$ and $\tau = 2 \phi$.  This metric is
that of $\IR^2 \times \IR^2$ provided that $\tau$ has period
$4\pi$ so that $\phi$ has period $2\pi$.  The $U(1)$ acts in one
of the $\IR^2$ planes and so this is the natural boundary
condition appropriate to a system with this symmetry.

Another possible boundary condition is is to require:
\begin{equation}
\nu ~\sim~ \log \bigg( {z^2 \over \big(1 + \coeff{1}{8}\, (x^2 +
y^2) \big)^2} \bigg) \,, \label{nuasympb}
\end{equation}
and then
\begin{equation}
ds^2_4 ~\sim~2 \, z^{-1} dz^2 ~+~  \coeff{1}{2}\, z\,
(d\tau+A_0)^2 ~+~ z\, {dx^2 ~+~ dy^2 \over  \big(1 +
\coeff{1}{8}\, (x^2 + y^2) \big)^2}     \,, \label{baseasympb}
\end{equation}
where
\begin{equation}
A_0 ~=~  \frac{1}{2}\,{(x dy ~-~ y dx) \over  \big(1 +
\coeff{1}{8}\, (x^2 + y^2) \big) }     \,. \label{Azero}
\end{equation}
Now set $x =\tan {\theta \over 2} \cos \phi $ and $y =\tan {\theta
\over 2} \sin \phi $ and one arrives at the metric:
\begin{eqnarray}
ds^2_4 &\sim& 2 \, z^{-1} dz^2 ~+~  \coeff{1}{2}\, z\,  (d\tau+
2\, (1-\cos \theta)
\, d\phi)^2 ~+~  2 \, z\,  \big(d\theta^2 ~+~ \sin^2 \theta d \phi^2 \big)  \nonumber\\
&\sim&   dr^2 ~+~  \coeff{1}{4}\,r^2\,  \big( \sigma_1^2 ~+~
\sigma_2^2 ~+~ \sigma_3^2 \big) \,, \label{baseasympc}
\end{eqnarray}
where $z = \frac{1}{8} r^2$,  the $\sigma_i$ are the left
invariant one-forms:
\begin{eqnarray}
\sigma_1 &\equiv&    \cos \psi \, d\theta   ~+~ \sin\psi \, \sin\theta \, d \phi \,, \nonumber\\
\sigma_2 &\equiv&    \sin\psi \, d\theta  ~-~ \cos\psi \, \sin\theta \, d \phi \,, \nonumber\\
\sigma_3 &\equiv &    d \psi ~+~ \cos \theta \, d \phi \,,
\label{sigmadefn}
\end{eqnarray}
and $ \tau= -2(\psi +\phi) $.  Once again, the $U(1)$ generated by
$\tau$ acts in one of the $\IR^2$ planes in $\IR^2 \times \IR^2 =
\IR^4$.

With either of these asymptotic behaviors, the  integral:
\begin{equation}
 \int \, \sqrt{\gamma}\, \gamma^{ij} \,  \partial_i \nu\, \partial_j  \nu  \, d^3 x    \,.
\label{volInt}
\end{equation}
converges at infinity.  The integrand is manifestly non-negative
and if $\nu$ is regular everywhere then we may integrate by parts.
This generates the Toda equation, (\ref{TodaEqn}), and so the
integral vanishes.  We therefore conclude that the only solution
 that is regular on $\IR^3$ is a constant.  Hence, $\nu$ must have singularities on $\IR^3$.

While general Toda metrics may have complicated singularities, we
are interested in metrics that, upon adding fluxes, give rise to
smooth bubbling solutions. For Gibbons-Hawking  base metrics, one
has positive and negative sources (GH points) for the metric
function, $V$, and pairs of these GH points then define the
homology cycles.  If one moves sufficiently close to one of these
singular points of $V$ in a GH metric, then the metric is, in
fact, regular and caps off into a piece of $\IR^4$ (perhaps
divided out by a discrete group) with $SO(4)$ rotation symmetry.
Guided by this, it is natural to consider singularities in $\nu$
that lead to local geometry that looks like $\IR^4/\ZZ_q$ for some
integer, $q$, and which locally has an $SO(4)$ invariance about
the singular point.

Equivalently,  near the singularities of $\nu$, the  Toda metric
has a $U(1) \times U(1) \subset SO(4)$ symmetry and so can be
mapped into a Gibbons-Hawking form. Thus the interesting class of
metrics for bubbling should be those that can be put into
Gibbons-Hawking form in the immediate vicinity of each singular
point of $\nu$.  The non-trivial part of the Toda solution then
relates to the transitions between these special regions.  One can
thus think of the Toda function as quilting together a collection
of GH pieces.

It is elementary to see from  the foregoing that, in the
neighborhood of a singular point of charge $\pm 1$, one must have:
\begin{equation}
 \nu ~\sim~ \log |z - \alpha| \,, \qquad   \pm(z - \alpha) > 0  \,.
\label{GHpoints}
\end{equation}
With these choices the metric becomes precisely that of $\IR^4$
and is positive or negative definite depending on the sign of the
charge.  By taking the $z \to 0$ limit in (\ref{nuasympb}) one can
also see that for a point of charge $+2$ one has $\nu ~\sim~2\,
\log |z - \alpha|$.  One can continue to higher charges via a
series expansion in $z$ but the geometry gets more complicated.
This is because a charge $q$ leads to a local geometry that is
$\IR^4/\ZZ_q$.  In GH spaces this discrete identification was
factored out of the $U(1)$ fiber, but in a general Toda geometry
it will be factored out of the base and so the geometry near the
singular points of $\nu$ will involve orbifold points in $\IR^3$.
It is therefore simpler to restrict to geometric charges of $\pm
1$ and take the view that other geometric charges can be obtained
via mergers of the more fundamental unit charges.

While we do not yet know how to progress beyond these simple
observations, we believe that similar considerations will apply to
bubbled geometries constructed from completely general ambi-polar,
hyper-K\"ahler metrics.  In the neighborhood of singular points of
the K\"ahler potential they will locally be of GH form and so one
might at least construct an approximate description as a quilt of
GH patches with transition functions.  Indeed, with such an
approximating metric one might be able to establish existence
theorems and perhaps even count moduli in the same manner that Yau
established the existence of Calabi-Yau metrics.

\section{The Atiyah-Hitchin metric}

The Atiyah-Hitchin metric has the form \cite{Atiyah:1985dv,
Bakas:1996gf}:
\begin{equation}
ds^2~=~ \coeff{1}{4}\, a^2b^2c^2\, d\eta^2~+~  \coeff{1}{4 }\, a^2
\, \sigma_1^2 ~+~ \coeff{1}{4}\,b^2 \, \sigma_2^2  ~+~
\coeff{1}{4}\, c^2 \,  \sigma_3^2\,, \label{AHform}
\end{equation}
where the $\sigma_i$ are defined in (\ref{sigmadefn}) and satisfy
 $d\sigma_i=\frac{1}{2}\epsilon_{ijk}\sigma_j\wedge\sigma_k$.
For (\ref{AHform}) to be hyper-K\"ahler, the functions $a(\eta)$,
$b(\eta)$ and $c(\eta)$ must satisfy:
\begin{eqnarray}
 \frac{\dot{a}}{a}~=~ \coeff{1}{2}\, \big( (b-c)^2 ~-~ a^2 \big)\, \\
 \frac{ \dot{b}}{b} ~=~\coeff{1}{2}\, \big( (c-a)^2 ~-~  b^2\big)\, \\
 \frac{ \dot{c}}{c} ~=~ \coeff{1}{2}\, \big((a-b)^2 ~-~  c^2  \big) \,,
\label{abcsystem}
\end{eqnarray}
where the dot denotes ${d \over d\eta}$.

\subsection{The standard solution}

This system of equations  may be mapped onto a Darboux-Halphen
system by introducing $w_1=bc$, $w_2=ac$ and $w_3=ab$.  One then
finds
\begin{equation}
\frac{d}{d\eta}\, (w_1+w_2) ~=~ -2w_1w_2, \qquad \frac{d}{d\eta}\,
(w_2+w_3) ~=~ -2w_2w_3,\qquad \frac{d}{d\eta}\,(w_1+w_3) ~=~
-2w_1w_3 \,. \label{DHsystem}
\end{equation}
To solve this system one first defines a new coordinate, $\theta$,
via
\begin{equation}
d\eta ~=~ \frac{d\theta}{u^2(\theta)} \,, \label{thetadefn}
\end{equation}
where $u$ is defined to be the solution of
\begin{equation}
\frac{d^2u}{d\theta^2} ~+~ \frac{u}{4\sin^2 \theta}  ~=~ 0 \,.
\label{udefn}
\end{equation}
One then finds that the solutions are given by
\cite{Atiyah:1985dv}:
\begin{eqnarray}
w_1 &=& -uu' ~-~  \coeff{1}{2}\,   u^2 \, \csc \theta\,, \nonumber \\
w_2 &=&  -uu' ~+~ \coeff{1}{2}\, u^2\cot \theta \,,  \nonumber \\
w_3 &=& -uu'  ~+~ \coeff{1}{2}\,   u^2 \, \csc \theta\ \,,
\label{DHsol}
\end{eqnarray}
where the prime denotes derivative with respect to $\theta$.

One can find the explicit solution for $u$ in terms of elliptic
functions:
\begin{equation}
u(\theta)= {c_1 \over \pi} \, \sqrt{\sin \theta}\,
K\big(\sin^2\coeff{\theta}{2}\big) ~+~ {c_2 \over \pi} \,
\sqrt{\sin \theta}\,  K\big(\cos^2\coeff{\theta}{2}\big)  \,,
 \label{generalu}
\end{equation}
where $c_1$ and $c_2$ are constants and
\begin{equation}
K(x^2) ~\equiv~  \int_0^{\pi/2} \, (1-x^2\sin^2\varphi)^{-1/2}\,
d\varphi \,. \label{ellK}
\end{equation}
A first order system for three functions like (\ref{DHsystem})
should involve three constants of integration.  These are
represented by $c_1$, $c_2$ and the trivial freedom to shift
$\eta$ by a constant.   However, in order to get a regular,
positive definite metric one must choose only one of the
non-trivial solutions, which is then canonically normalized to:
\begin{equation}
u(\theta)= \frac{1}{\pi}\, \sqrt{\sin \theta}\,
K\big(\sin^2\coeff{\theta}{2}\big) \,. \label{canonu}
\end{equation}
With this choice, the function $u(\theta)$ is non-vanishing on
$(0,\pi)$ and so the change of variables (\ref{thetadefn}) is
well-defined.  Moreover one has $w_1 <0$, $w_2 <0$ and $w_3 >0$ on
$(0,\pi)$ and so the metric coefficients:
\begin{equation}
a^2  ~=~  {w_2 \, w_3 \over w_1} \,, \qquad b^2 ~=~ {w_1 \, w_3
\over w_2} \,, \qquad  c^2  ~=~ {w_1 \, w_2 \over w_3} \,,
\label{metcoeff}
\end{equation}
are all positive.

\subsection{The geometry of the Atiyah-Hitchin metric}

The standard Atiyah-Hitchin geometry is asymptotic to $\IR^3
\times S^1$ and has a non-trivial two-cycle, or ``bolt'' in the
center.   To see this we first look at the structure at infinity,
which corresponds to $\theta \to \pi$. In this limit one has:
\begin{eqnarray}
u(\theta)&\sim& - \coeff{1}{\pi} \,  \sqrt{2\cos
\coeff{\theta}{2}}\, \log(\cos \coeff{\theta}{2}) \,, \qquad d\eta
~\sim~ \frac{\pi^2 \,
d\theta}{2\cos\coeff{\theta}{2}(\log(\cos\coeff{\theta}{2}))^2}
\,,
\nonumber\\
w_1(\theta) &\sim& \coeff{1}{\pi^2} \, \log(\cos\coeff{\theta}{2})
\,, \quad w_2(\theta)  ~\sim~  \coeff{1}{\pi^2} \,
\log(\cos\coeff{\theta}{2}) \,, \quad w_ 3 (\theta) ~\sim~
\coeff{1}{\pi^2} \, (\log(\cos\coeff{\theta}{2}))^2  \,,
\label{wjasymp}
\end{eqnarray}
which implies
\begin{equation}
a(\theta)  ~\sim~   \coeff{1}{\pi} \, \log(\cos\coeff{\theta}{2})
\,, \qquad b(\theta) ~\sim~ \coeff{1}{\pi} \,
\log(\cos\coeff{\theta}{2}) \,, \qquad c(\theta) ~\sim~
\coeff{1}{\pi} \,. \label{metasymp}
\end{equation}
Define  $r=-\log(\cos\frac{\theta}{2})$ and then the asymptotic
form of the metric becomes:
 \begin{equation}
ds^2~\sim~\coeff{1}{4\, \pi^2}\, \big( dr^2 ~+~ r^2(\sigma_1^2 ~+~
\sigma_2^2)  ~+~\sigma_3^2 \big) \,,
\end{equation}
which indeed has the structure of a $U(1)$ fibration over $\IR^3$.

At the other end of the interval, $\theta \to 0$, one finds:
\begin{eqnarray}
u(\theta)&\sim& \coeff{1}{2}\,  \theta^{1\over 2} ~-~
\coeff{1}{96}\, \theta^{5\over 2} ~+~
\cO(\theta^{{7\over 2}})\,,  \qquad  d\eta ~\sim~  4 \theta^{-1}  d\theta  \nonumber \\
w_1(\theta)&\sim& -\coeff{1}{4} ~-~ \coeff{1}{2048}\,\theta^{4}
~+~ \cO(\theta^{6}) \,, \qquad w_2(\theta) ~\sim~ -\coeff{1}{32}\,
\theta^2 ~-~ \coeff{1}{3072}\, \theta^{4} ~+~ \cO(\theta^{6})\,,
 \nonumber \\
w_3(\theta) &\sim& \coeff{1}{32}\, \theta^2  ~+~ \coeff{7}{3072}\,
\theta^{4}  ~+~ \cO(\theta^{6})\,, \qquad a(\theta)~\sim~
\coeff{1}{16}\, \theta^2~+~ \coeff{1}{384}\, \theta^4 ~+~
\cO(\theta^{6}),
\nonumber \\
b(\theta) &\sim& \coeff{1}{2}  ~+~ \coeff{1}{64}\,\theta^{2}  ~+~
\cO(\theta^{4})\,, \qquad
 \qquad c(\theta)~\sim~ \coeff{1}{2} ~-~ \coeff{1}{64}\,\theta^{2}  ~+~ \cO(\theta^{4})
\end{eqnarray}
Define $\rho = {1 \over 64} \theta^2$ and the metric near $\theta
=0$ has the form:
\begin{equation}
ds^2~\sim~ d\rho^2 ~+~4\, \rho^2 \,  \sigma_1^2 ~+~
\coeff{1}{16}\, \big(\sigma_2^2~+~ \sigma_3^2\big)
\end{equation}
Thus we see the ``bolt''  at the origin.  Note that the scale of
the metric has been fixed and the radius of the bolt has been set
to $\frac{1}{4}$. The fact that the coefficient of $\sigma_1$
vanishes as $\sim 4 \rho^2$ also has important implications for
the global geometry.  There is a very nice discussion of this in
the appendices of \cite{Cvetic:2001sr}.

For future reference, we will chose the constant of integration
(\ref{thetadefn}) so that $\eta \to 0$ at infinity ($\theta =
\pi$) and take:
\begin{equation}
\eta ~\equiv~    - \int^\pi_\theta  \, {d \theta \over u^2} \,.
\label{etadefn}
\end{equation}
With this choice, $\eta$ has the following asymptotic behavior:
\begin{equation}
\eta ~\sim~  4\, \log(\theta)  \ \ {\rm as} \ \  \theta \to 0\,;
\qquad\qquad
 \eta ~\sim~  - {\pi^2 \over r}   \ \ {\rm as} \ \    \theta \to \pi\,,
\label{etaasymp}
\end{equation}
where $r=-\log(\cos\frac{\theta}{2})$.

Since there is a non-trivial two-cycle, there must be a
non-trivial, dual element of cohomology.  That is, there must be
precisely one square-integrable, harmonic two-form.   In
particular, this means the two-form must be a singlet under
$SO(3)$.  To determine this two form, it is convenient to
introduce the vierbeins:
\begin{equation}
e^1~=~ - \coeff{1}{2}\,  abc \, d\eta\,, \qquad\qquad
e^2~=~\coeff{1}{2}\,  a \, \sigma_1\,,
 \qquad e^3~=~ \coeff{1}{2}\,  b \, \sigma_2\,, \qquad\qquad
e^4~=~ \coeff{1}{2}\,  c \, \sigma_3\,, \label{AHframes}
\end{equation}
and define some manifestly $SO(3)$-invariant, self-dual two-forms
via:
\begin{eqnarray}
\Omega_1 &\equiv& h_1\,\big( a^2\, d\eta\wedge\sigma_1~-~
\sigma_2\wedge \sigma_3\big)\,,  \nonumber\\
\Omega_2 &\equiv& h_2\,\big(b^2\, d\eta\wedge\sigma_2 ~+~
\sigma_1\wedge \sigma_3\big) \,, \nonumber\\
\Omega_3 &\equiv& h_3\,\big(c^2\, d\eta\wedge\sigma_3 ~-~
\sigma_1\wedge \sigma_2\big)\,, \label{AHforms}
\end{eqnarray}
for some functions, $h_j(\eta)$.  The condition that $\Omega_j$ be
closed, and hence harmonic is:
\begin{equation}
{d \over d\eta}\, \log( h_j)  ~=~ - a_i^2 \quad \Leftrightarrow
\quad {d \over d \theta}\, \log( h_j)  ~=~ - {a_i^2  \over u^2}
\,, \label{hjeqn}
\end{equation}
where
\begin{equation}
(a_1,a_2,a_3) \equiv (a,b,c)\,.
\end{equation}
These equations imply that there are obvious local potentials,
$B_j$, for $\Omega_j$:
\begin{equation}
\Omega_j ~=~ d \, B_j \,, \quad{\rm where} \quad B_j ~\equiv~ -
h_j\, \sigma_j \,. \label{Bjdefn}
\end{equation}
Remarkably enough, the equations for the $h_j$ are integrable in
terms of $u(\theta)$ and we find:
\begin{equation}
h_1  ~=~\coeff{1}{4}\, \alpha_1\, { u^2 \over w_1 \, \sin({\theta
\over 2})} \,, \qquad h_2  ~=~ \coeff{1}{4}\, \alpha_2\, { u^2
\over w_2} \,, \qquad h_3  ~=~  \coeff{1}{4}\, \alpha_3\,{  u^2
\over w_3 \, \cos({\theta \over 2})} \,, \qquad \label{hjres}
\end{equation}
where the $\alpha_j$ are constants of integration.  One should
note that these solutions follow from (\ref{udefn}) and
(\ref{DHsol}) and do not depend upon the specific choice in
(\ref{canonu}).   However here we focus on the solutions that
arise from (\ref{canonu}). To determine which, if any, of the
$h_j$ gives the desired harmonic form, we look at the regularity
of these two-forms and examine their behavior as $\theta \to 0$
and $\theta \to \pi$.

As $\theta \to 0$ we have:
\begin{equation}
h_1  ~\sim~ -\coeff{1}{2}\, \alpha_1 ~+~ \cO(\theta^4 ) \,, \qquad
h_2  ~\sim~ - 2\, \alpha_2  \, \theta^{-1} ~+~ \cO(\theta) \,,
\qquad h_3  ~\sim~  2\, \alpha_3  \, \theta^{-1} ~+~ \cO(\theta)
\,, \label{hjzero}
\end{equation}
and as $\theta \to \pi$ we have:
\begin{equation}
h_1  ~\sim~ \coeff{1}{4}\, \alpha_1  \, r\, e^{-r}  ~+~ \cO(e^{-r}
) \,, \qquad h_2  ~\sim~\coeff{1}{4}\, \alpha_2  \,  r\, e^{-r}
~+~ \cO(e^{-r} )  \,, \qquad h_3  ~\sim~  \coeff{1}{2}\, \alpha_3
~+~  O(r^{-1} )   \,, \label{hjpi}
\end{equation}
where $r=-\log(\cos\frac{\theta}{2})$.  It follows that $h_1$ is
regular at $\theta =0$ and falls off very fast at infinity.   The
corresponding two-form, $\Omega_1$,  is globally regular and
square-integrable and is thus the harmonic form we seek.   Indeed,
at $\theta =0$ one has  $\Omega_1 = {1 \over 2}\alpha_1 \sigma_2
\wedge \sigma_3$ and $\sigma_2 \wedge \sigma_3$ is the volume form
on the bolt of unit radius, which means the period integral is
given by:
\begin{equation}
\int_{Bolt} \, \Omega_1 ~=~ \coeff{1}{2}\, \alpha_1 \int_{Bolt} \,
\sigma_2 \wedge \sigma_3  ~=~ 2\, \pi \, \alpha_1   \,.
\label{fluxint}
\end{equation}
%

\subsection{Ambi-polar Atiyah-Hitchin metrics }
 \label{APAH}

The most general $SO(3)$ invariant metric governed by
(\ref{DHsystem}) requires one to use the most general function,
$u(\theta)$,  in  (\ref{generalu}). As we will see, this
possibility is usually ignored because it leads to ambi-polar
metrics, and we will show, in the next section, how such solutions
can be used to make new Lorentzian BPS solutions in five
dimensions.

To understand how the inclusion of the extra function changes the
Atiyah-Hitchin metric, define $\tilde u(\theta) \equiv  u(\pi
-\theta) $ and let  $\tilde  w_j (\theta)$ be defined by
(\ref{DHsol}) with $u$ replaced by $\tilde u$.   It is evident
that $\tilde u(\theta)$ also solves (\ref{udefn}), indeed, it
simply interchanges $c_1$ and $c_2$ in (\ref{generalu}).
Therefore the functions $\tilde w_j$ also solve the system
(\ref{DHsystem}).  On the other hand, from (\ref{DHsol}) one can
easily see that:
\begin{equation}
w_1 (\pi -\theta) ~=~  -\tilde  w_3( \theta) \,, \qquad  w_2 (\pi
- \theta) ~=~ -\tilde  w_2(\theta) \,, \qquad  w_3 (\pi -  \theta)
~=~  - \tilde w_1( \theta) \,. \label{invertwj}
\end{equation}
Thus allowing a non-zero value for $c_1$ and $c_2$ means that
asymptotic behavior of the $w_j$ at $\theta =0$ is related to the
asymptotic behavior at $\theta =\pi$.  In particular,  because we
now have
\begin{eqnarray}
u(\theta) &\sim&  - \coeff{c_1}{\pi} \,  \sqrt{2\cos
\coeff{\theta}{2}}\, \log(\cos \coeff{\theta}{2}) \,,
\qquad \theta \to \pi\,, \nonumber \\
u(\theta) &\sim& - \coeff{c_2}{\pi} \,  \sqrt{2\sin
\coeff{\theta}{2}}\, \log(\sin \coeff{\theta}{2}) \,, \qquad
\theta \to 0\,, \label{genuasymp}
\end{eqnarray}
we therefore have, as $\theta \to \pi$:
\begin{equation}
w_1(\theta) ~\sim~ \coeff{c_1^2}{\pi^2} \,
\log(\cos\coeff{\theta}{2}) \,, \quad w_2(\theta)  ~\sim~
\coeff{c_1^2}{\pi^2} \, \log(\cos\coeff{\theta}{2}) \,, \quad w_ 3
(\theta) ~\sim~  \coeff{c_1^2}{\pi^2} \,
(\log(\cos\coeff{\theta}{2}))^2  \,, \label{genwjasympa}
\end{equation}
and, as $\theta \to 0$:
\begin{equation}
w_1(\theta) ~\sim~ -\coeff{c_2^2}{\pi^2} \,
(\log(\sin\coeff{\theta}{2}))^2 \,, \quad w_2(\theta)  ~\sim~  -
\coeff{c_2^2}{\pi^2} \, \log(\sin\coeff{\theta}{2}) \,, \quad w_ 3
(\theta) ~\sim~   - \coeff{c_2^2}{\pi^2} \,
\log(\sin\coeff{\theta}{2}) \,. \label{genwjasympb}
\end{equation}
This means that the metric now has two regions that are asymptotic
to $\IR^3 \times S^1$ with $a \sim r$ and $b \sim r$ as $\theta
\to \pi$ and with $c \sim r$ and $b \sim r$ as $\theta \to 0$.  It
therefore, naively looks like a ``wormhole'' geometry. The
asymptotics also imply that if the metric is positive definite in
one asymptotic region then it is negative definite in the other:
$a^2, b^2$ and $c^2$ all change sign as one goes from $\theta =0$
to $\theta = \pi$.  One also sees from the asymptotics of $w_2$
that $w_2$ must have at least one zero in $(0,\pi)$ and so the
metric is singular at such a point.   It is for all these reasons
that the generalization of the Atiyah-Hitchin metric is usually
ignored. However, this metric is ambi-polar and, as we will show,
all the pathologies itemized here are not present in the
five-dimensional solution that can be constructed from this
metric.

\begin{figure}
\centering
\includegraphics[height=6cm]{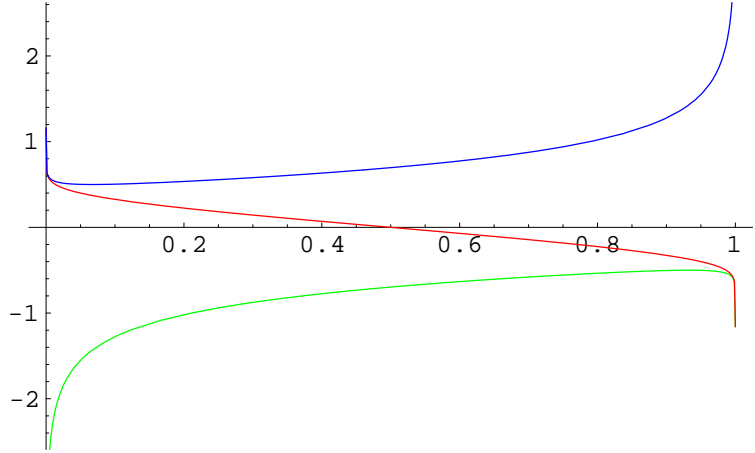}
\caption{This shows the three functions, $w_j$, as a function of
$x = \sin^2\coeff{\theta}{2}$ when $u$ is given by
(\ref{simpgenu}).  One has $w_1 < 0$ and $w_3 > 0$ and $w_2$ has a
simple zero at $\theta = \pi/2$.  All three functions diverge at
both ends of the interval.}
\label{wjplots}
\end{figure}

For simplicity, we will restrict our attention in this paper to
ambi-polar metrics based upon:
\begin{equation}
u(\theta)= {1 \over \pi} \, \sqrt{\sin \theta}\, \big(
K\big(\sin^2\coeff{\theta}{2}\big) ~+~
  K\big(\cos^2\coeff{\theta}{2}\big) \big)  \,,
 \label{simpgenu}
\end{equation}
then one has
\begin{eqnarray}
w_1 (\pi -\theta) &=& -  w_3( \theta) \,, \qquad  w_2 (\pi -
\theta) ~=~
- w_2(\theta) \,, \qquad  w_3 (\pi -  \theta) ~=~  -  w_1( \theta) \,, \nonumber\\
a^2 (\pi -\theta) &=& -  c^2 ( \theta) \,, \qquad  b^2 (\pi -
\theta) ~=~ - b^2 (\theta) \,, \qquad  c^2 (\pi -  \theta) ~=~  -
a^2( \theta) \,. \label{reflectw}
\end{eqnarray}
With this choice one has $u>0$, $w_1 < 0$ and $w_3 > 0$  for
$\theta \in [0,\pi]$ and $w_2$ has a simple zero at $\theta =
\pi/2$.  See Fig. \ref{wjplots}. This means that the metric
coefficients, $a_j^2$, simultaneously change sign at $\theta =
\pi/2$ and this is the only point at which this happens.
Moreover, $a^2$ and $c^2$ have simple zeroes while $b^2$ has a
simple pole at $\theta = \pi/2$.  This behavior of the metric
coefficients precisely mimics that of the ambi-polar GH metrics.

We note that the forms given by (\ref{AHforms}) and (\ref{hjres})
are still ``harmonic'' in that they are self-dual and closed.
Moreover, $\Omega_1$ and $\Omega_3$ are non-singular in the
wormhole geometry, except that $\Omega_1$ remains finite as
$\theta \to 0$ while $\Omega_3$ remains finite as $\theta \to
\pi$.  This means that neither is square-integrable on the
complete geometry.  On the other hand, $\Omega_2$ falls off
exponentially at both $\theta = 0$ and $\theta = \pi$ but is
singular at $\theta = \pi/2$, where the metric changes sign.  Once
again this last flux has a behavior precisely analogous to the
two-form fields that were essential building blocks for the
regular five-dimensional solutions that can be built from
ambi-polar GH metrics.

Finally, we should comment that more general choices of
$u(\theta)$, such as taking $c_1 = - c_2 =1$ in (\ref{generalu}),
can result in solutions with zeroes for $w_1$, $w_2$ and $w_3$.
We have not studied these in detail.

\section{The BPS solutions}

\subsection{Solving the BPS equations}
\label{StandardSol}

Since there is only one independent harmonic form in the
Atiyah-Hitchin metric, this means that the two-forms,
$\Theta^{(I)}$, in (\ref{BPSeqn:1}) must all be proportional to
one another for $I=1,2,3$.  For simplicity, we will, in fact, take
them all to be equal. We will also take the three warp factor
functions to be equal, $Z_I = Z$, $I=1,2,3$. Ignoring, for the
present, issues of regularity, the $SO(3)$ invariant solutions of
(\ref{BPSeqn:1}) are given by the $\Omega_i$ of (\ref{AHforms})
and so we will take
\begin{equation}
\Theta^{(I)}~=~  \Theta ~=~  \Omega_1 ~+~ \Omega_2 ~+~ \Omega_3\,.
 \label{ThetaAH}
\end{equation}
The functions, $h_j$, in (\ref{hjres}) contain integration
constants, $\alpha_j$, that make this an arbitrary linear
combination.  Note:  One should not confuse the index, $I= 1,2,3$
on $\Theta^{(I)}$ with the index, $i=1,2,3$ on $\Omega_i$. The
former indexes the $U(1)$ gauge groups of three-charge system
while the latter labels the three distinct type of two-form in
(\ref{AHforms}) that satisfy (\ref{BPSeqn:1}).

With this choice, the  second BPS equation becomes:
\begin{equation}
{d^2 Z\over d\eta^2}    ~=~  8\, \sum_{j=1}^3 \,h_j^2 \, a_j^2\,.
 \label{BPSAH:2}
\end{equation}
Given the form of $\Theta$, there is a unique Ansatz for the
angular momentum vector, $k$:
\begin{equation}
k  ~=~    \sum_{j=1}^3 \,\mu_j  \, \sigma_j\,,
 \label{muAHform}
\end{equation}
which means that the third BPS equation yields three equations:
\begin{equation}
{d  \mu_j\over d\eta }\,  ~-~ a_j^2 \, \mu_j  ~=~  3\,  h_j  \,
a_j^2 \, Z\,,  \qquad {j=1,2,3}\,.
 \label{BPSAH:3}
\end{equation}
The factor of three comes from the sum over the $U(1)$ label, $I$,
in (\ref{BPSeqn:3}) and the choices: $\Theta^{(I)}= \Theta$, $Z_I
= Z$.

These equations can, once again, be integrated explicitly in terms
of the the elliptic function, $u$.   First, from (\ref{hjeqn}) we
have:
\begin{equation}
{d Z\over d\eta }    ~=~  \gamma_0 ~-~  4\, \sum_{j=1}^3 \,h_j^2
\,,
 \label{Zint:1}
\end{equation}
for some constant, $\gamma_0$.  Using (\ref{udefn}) and
(\ref{DHsol})  one can easily show that
\begin{equation}
{d  \over d \eta}\, {\alpha_j^2 \over w_j}  ~=~ u^2 \, {d  \over d
\theta}\, {\alpha_j^2  \over w_j}   ~=~      \alpha_j^2  ~+~ 4 \,
(-1)^j\,h_j^2 \,, \label{derwj}
\end{equation}
and hence:
\begin{equation}
Z    ~=~  \delta ~+~ \gamma \, \eta ~-~ \sum_{j=1}^3 \, (-1)^j\,
{\alpha_j^2  \over w_j}   \,,
 \label{Zint:2}
\end{equation}
where  $\gamma =  \gamma_0  + \sum_{j=1}^3   (-1)^j \alpha_j^2$.

The last BPS equation, (\ref{BPSAH:3}),  can be integrated to
yield:
\begin{equation}
\mu_j  ~=~  {3 \over h_j}\,  \int h_j^2  \, a_j^2 \, Z \, d\eta
~=~  {3 \over h_j}\,  \int \Big( - {1 \over 2}\, {d \over d\eta}
h_j^2 \Big) \, Z \, d\eta  \,,  \qquad {j=1,2,3}\,.
 \label{muAHsol}
\end{equation}
It is easy to integrate this explicitly.  First, by integrating by
parts one can show:
\begin{equation}
 {3 \over h_j}\,  \int h_j^2  \, a_j^2 \, \big( \delta ~+~ \gamma \, \eta ) \, d\eta  ~=~
-\coeff{3}{2}\, \delta\, h_j ~-~ \coeff{3}{2}\, \gamma\,\Big[ \,
h_j \eta  ~-~ (-1)^j \,
  {\alpha_j^2 \over 4\, h_j } \, \Big({1 \over w_j}  - \eta\Big)\Big]
  ~+~ {\beta_j  \over   h_j }   \,,
\label{mulinpart}
\end{equation}
where the $\beta_j$ are constants of integration.  The other parts
of the integrals for $\mu_j$ can be obtained from:
\begin{eqnarray}
{3 \over h_j}\,  \int \, {h_j^2  \, a_j^2\over w_j} \,  d\eta  &=&
(-1)^j\, {\alpha_j^2 \over 8\, h_j}\,
\Big[ \,{2\, w_i \, w_k \over w_j^3}  ~-~ {w_i +  w_k \over w_j^2} \, \Big]\,, \nonumber\\
{3 \over h_j}\,  \int \, {h_j^2  \, a_j^2\over w_i} \,  d\eta  &=&
(-1)^{j+1}\, {3\, \alpha_j^2 \over 8\, h_j}\,  {(w_j -  w_k) \over
w_j^2} \,, \label{muotherpart}
\end{eqnarray}
where $i,j,k \in \{1,2,3\}$ are all distinct.

Thus, rather surprisingly, we can obtain the complete solution
analytically in terms of elliptic functions.

\subsection{The bubbled solution on the standard Atiyah-Hitchin base}

The  physical intuition underlying BPS solutions is that all
charges have to be of the same sign so that the electromagnetic
repulsion balances the gravitational attraction.   Bubbled
geometries generically have geometric charges of all signs and
then the attractive forces are balanced by threading cycles with
fluxes that then resist the collapse of the bubbles.  The result
is then a stable configuration where the sizes of some of the
bubbles are fixed in terms of the fluxes that thread them.  Such
relationships are typically embodied in a system of  ``Bubble
Equations''  \cite{Bena:2005va,Berglund:2005vb}. If one insists
that a solution is a BPS configuration but one does not have the
forces properly balanced then the solution is then supported only
through the appearance of CTC's.  Thus, when investigating BPS
geometries one typically encounters the constraints of bubble
equations through the process of eliminating CTC's.

The standard Atiyah-Hitchin base metric is, in its own right, a
well-behaved BPS solution with no additional fluxes.  Indeed, the
addition of a flux through the non-trivial two-cycle should drive
the configuration out of equilibrium and expand the bubble.  We
should therefore find irremovable CTC's if we attempt to include a
non-trivial flux.   We now show that this is precisely what
happens.

As we remarked earlier, the only non-trivial, harmonic flux on the
standard Atiyah-Hitchin base is given by $\Omega_1$ and so we set
$\alpha_2 = \alpha_3 =0$ in the results of the previous
sub-section\footnote{If one is interested in
  solutions that are asymptotically $AdS \times S^2$, one could also
  investigate solutions that contain the $\Omega_3$ component of the
  2-form field strength $\Theta$, which corresponds to constant flux on the $S^2$.
  Nevertheless, in our investigations this did not give any sensible solutions.
}.  We then find:
\begin{equation}
Z    ~=~  \delta ~+~ \gamma \, \eta ~+~   {\alpha_1^2  \over w_1}
 \label{ZregAH}
\end{equation}
and $k = \mu \sigma_1$,  where
\begin{eqnarray}
\mu  &=&   -\coeff{3}{2}\, \delta\, h_1 ~-~ \coeff{3}{2}\,
\gamma\, \Big[ \, h_1 \eta  ~+~  \,  {\alpha_1^2 \over 4\, h_1 }
\,
\Big({1 \over w_1}  - \eta\Big)\Big]  \nonumber \\
&& ~-~ \, {\alpha_1^4 \over 8\, h_1}\,  \Big[ \,{2\, w_2 \, w_3
\over w_1^3}  ~-~ {w_2 +  w_3 \over w_1^2} \, \Big] ~+~ {\beta_1
\over   h_1 }   \,.
 \label{muegAH}
\end{eqnarray}
\\
It is interesting to note that the part of $Z$ corresponding to
the flux sources in (\ref{ZregAH}) ({\it i.e.} the $w_1^{-1}$
term) is always negative, and therefore at infinity this warp
factor looks like it is coming from an object of negative mass and
charge. This is however not surprising, considering that the
Atiyah-Hitchin space also looks asymptotically as a negative-mass
Taub-NUT space.

The value of $\beta_1$ is fixed by requiring that $\mu$ does not
diverge, and indeed falls off at infinity.  We find that if we
set:
\begin{equation}
\beta_1 ~=~  \frac{\pi^ 2\,\alpha_1^4 }{8} \,,
 \label{betafix:1}
\end{equation}
then this removes all the terms that diverge at infinity and
leaves only terms that fall off. Indeed, there are two types of
such terms:  Those proportional to $\gamma$, which fall off as ${1
\over r}$, and the remainder that fall off as  $r e^{-r}$.

Near $\theta =0$ the function $\eta$ is logarithmically divergent
and so $Z$ is logarithmically divergent unless $\gamma =0$.
Physically, a non-zero value of $\gamma$ corresponds to a uniform
distribution of M2 branes smeared over the bolt at $\theta =0$,
with negative values of $\gamma$ corresponding to positive charge
densities.    If $\gamma =0$  then $ Z = \delta - 4\alpha_1^2$ at
$\theta =0$.

For constant time slices, the five-dimensional metric
(\ref{fivemetric}) becomes
\begin{equation}
ds^2 ~=~  \big(\coeff{1}{4} \, a^2\, Z ~-~  \mu^2\, Z^{-2} \big)\,
\sigma_1^2 ~+~
 \coeff{1}{4} \, Z \,a^2\, b^2\, c^2 \, d\eta^2 ~+~ \coeff{1}{4} \,Z\, b^2\sigma_2^2 ~+~ \coeff{1}{4} \,
Z\, c^2\sigma_3^2 \,, \label{ssmetric}
\end{equation}
and so to avoid CTC's, one must have $Z\ge 0$ and the quantity:
\begin{equation}
\cQ~=~   \coeff{1}{4} \, a^2\, Z^3 ~-~  \mu^2 \, \label{cQdefn}
\end{equation}
must be non-negative.  The function $a(\theta) \sim \frac{1}{16}
\theta^2$ as $\theta \to 0$ and $Z$ diverges, at worst,
logarithmically.  Thus we must have $\mu \to 0$ as $\theta \to 0$
in order to avoid CTC's on the bolt.  (This is how the bubble
equations arise on GH spaces.)  This means  that we must take
\begin{equation}
\gamma ~=~   \coeff{1}{4} \, \delta ~-~  \coeff{2}{3} \,
\alpha_1^{-2} \, \beta_1 ~=~\coeff{1}{4} \, \delta ~-~
\coeff{1}{12} \,  \pi^ 2 \, \alpha_1^2  \,. \label{gamres}
\end{equation}

For pure-flux solutions, which have no singular sources, one must
take  $\gamma=0$ and the CTC condition (\ref{gamres}) reduces to
$\delta =  \frac{1}{3} \,  \pi^ 2 \, \alpha_1^2$. Then one finds
\begin{equation}
 \coeff{1}{4} \, a^2\, Z ~-~  \mu^2\, Z^{-2}   ~\sim~ - \coeff{1}{3072} \,(12 -\pi^2) \,
 \alpha_1^2 \, \theta^4 ~<~ 0\,,
\label{CTCprob}
\end{equation}
and so one necessarily has CTC's in the immediate neighborhood of
the bolt. This is a signal that there is no physical BPS solution
based upon the standard Atiyah-Hitchin metric with pure flux: The
flux will blow up the cycle and there is no gravitational
attraction holding the bubble back.

\begin{figure}
\centering
\includegraphics[height=6cm]{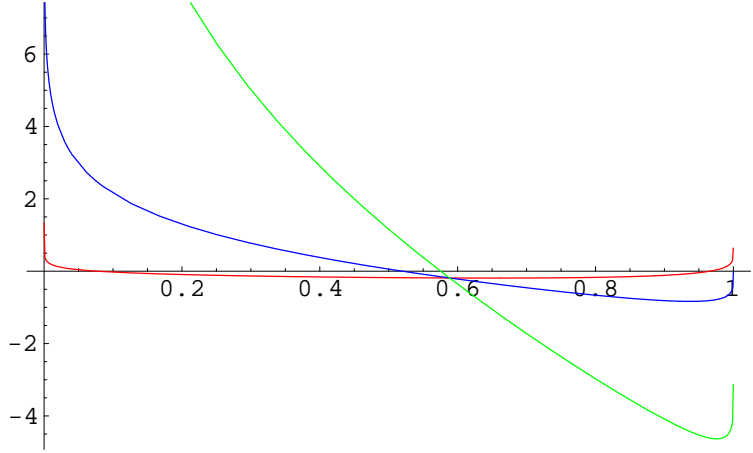}
\caption{This shows plots of $Z$ as a  function of $x =
\sin^2\coeff{\theta}{2}$. We have taken $\delta =1$, fixed
$\gamma$ in terms of $\alpha_1$ using (\ref{gamres}) and then we
have chosen three values of $\alpha_1$ that ensure that $\gamma$
is negative: $\alpha_1= 0.6$, $\alpha_1= 1.0$ and $\alpha_1=2.0$.
The steeper graphs  at $x=0.5$ correspond to larger values of
$\alpha_1$.  Note that $Z \to 1$ as $x \to 1$, but that $Z$ is
generically negative for $x>0.6$.}
\label{badZ}
\end{figure}

One might hope that one could stabilize the solution with a
distribution of M2 branes on the bolt.  While this might be
possible in general, it does not seem to be possible with a
uniform, $SO(3)$ invariant distribution. For this, one must have
$\gamma <0$ for $Z$ to remain positive near $\theta =0$ and then
(\ref{gamres}) means that $\alpha_1^2 > {3 \over \pi^2} \delta$.
In addition, we must have $\delta \ge 0$ for $Z>0$ at infinity.
From  (\ref{Zint:1}) one has
\begin{equation}
{d Z\over d\eta }    ~=~  \gamma~+~ \alpha_1^2 ~-~  4  \,h_1^2 ~=~
\coeff{1}{4} \, \delta ~+~ (1 -  \coeff{1}{12} \,  \pi^ 2 ) \,
\alpha_1^2  ~-~ 4  \,h_1^2    \,,
 \label{derZ}
\end{equation}
and since  $h_1 = -\frac{1}{2}\, \alpha_1$ at $\theta =0$  and
$h_1  \to 0$ at infinity ($\theta =\pi$)  we see that ${d Z\over
d\eta }$ is negative at $\theta =0$ and positive at $\theta =\pi$.
Therefore, $Z$ has a minimum for $\theta \in (0,\pi)$.  While we
have not done an exhaustive analysis, we generally find that $Z$
is negative at this minimum value.  Some examples are shown in
Fig. \ref{badZ}.   Obviously, the complete five-dimensional
metric is singular when $Z <0$.

Adding the singular M2-brane sources does render $\cQ$ positive in
a region around the bolt but, as one can see from (\ref{cQdefn}),
$\cQ$ also goes negative  shortly before $Z$ goes negative.  Thus
adding M2 branes sources moves CTC's away from the bolt but at the
cost of more extensive singular behavior elsewhere in the
solution.

\subsection{Bubbling the ambi-polar Atiyah-Hitchin base}

We now consider adding flux  to one of the ambi-polar
Atiyah-Hitchin metrics discussed in Section  \ref{APAH}.  That is,
we will start with the ambi-polar ``wormhole'' geometry that
arises from the choice  (\ref{simpgenu}), which therefore has the
reflection symmetry given by (\ref{reflectw}).  The solutions to
the BPS equations have exactly the same functional form as those
given in Section \ref{StandardSol} for the standard Atiyah-Hitchin
background. However, the underlying functions now have very
different asymptotic behavior and this affects all of the choices
based upon regularity and square integrability.

Let $r = -\log(\pi - \theta)$ and $\hat r = -\log( \theta)$, then
as $\theta \to 0$ one has
\begin{eqnarray}
u(\theta)&\sim&  \coeff{1}{\pi} \, \hat r\,  e^{-{\hat r/2}}  \,,
\qquad \eta ~\sim~ - \eta_0~+~  \frac{\pi^2 }{ \hat r} \,,
\nonumber\\
w_1(\theta) &\sim&  -\coeff{1}{\pi^2} \, \hat r^2  \,, \quad
w_2(\theta)  ~\sim~   \coeff{1}{\pi^2} \, \hat r  \,, \quad w_ 3
(\theta) ~\sim~ \coeff{1}{\pi^2} \, \hat r   \,,
\label{APwjasymp:1}
\end{eqnarray}
which implies
\begin{equation}
a^2(\theta) ~\sim~   - \coeff{1}{\pi^2} \,, \qquad b^2(\theta)
~\sim~ -\coeff{1}{\pi^2} \, \hat r^2 \,, \qquad c^2(\theta) ~\sim~
-\coeff{1}{\pi^2} \, \hat r^2\,. \label{APmetasymp:1}
\end{equation}
The constant, $\eta_0$, is defined by\footnote{While we haven't
proven that $\eta_0 = 2\pi$ analytically, we have checked
numerically to over $100$ significant figures.}:
\begin{equation}
\eta_0 ~\equiv~  \int_0^\pi \, {1 \over u(\theta)^2} ~=~2\pi \,.
\label{etazerodefn}
\end{equation}
As $\theta \to \pi$ one has:
\begin{eqnarray}
u(\theta)&\sim& \coeff{1}{\pi} \,  r \,  e^{-{  r/2}}  \,, \qquad
\eta ~\sim~ - \frac{\pi^2}{ r}  \,,
\nonumber\\
w_1(\theta) &\sim& - \coeff{1}{\pi^2} \,  r   \,, \quad
w_2(\theta)  ~\sim~ - \coeff{1}{\pi^2} \,  r  \,, \quad w_3
(\theta) ~\sim~  \coeff{1}{\pi^2} \,  r^2  \,, \label{APwjasymp:2}
\end{eqnarray}
which implies
\begin{equation}
a^2(\theta)  ~\sim~  \coeff{1}{\pi^2} \, r^2 \,, \qquad
b^2(\theta) ~\sim~  \coeff{1}{\pi^2} \,  r^2 \,, \qquad
c^2(\theta) ~\sim~ \coeff{1}{\pi} \,. \label{APmetasymp:2}
\end{equation}
The metric in each of these asymptotic regions becomes:
 \begin{eqnarray}
ds^2 &\sim& - \coeff{1}{4\, \pi^2}\, \big( d \hat r^2 ~+~  \hat
r^2(\sigma_2^2 ~+~ \sigma_3^2)  ~+~
\sigma_1^2 \big) \,, \qquad \theta \to 0 \,,  \nonumber \\
ds^2 &\sim& \coeff{1}{4\, \pi^2}\, \big( dr^2 ~+~  r^2(\sigma_1^2
~+~ \sigma_2^2)  ~+~ \sigma_3^2 \big) \,, \qquad \theta \to \pi
\,. \label{APasympmet}
\end{eqnarray}
We thus have an ambi-polar metric with two regions that are
asymptotic to different $U(1)$ fibrations over different $\IR^3$
bases.  The metric changes sign precisely at $\theta
=\frac{\pi}{2}$ at which point the metric function $b^2(\theta)$
has a simple pole, while $a^2(\theta)$ and $c^2(\theta)$ have
simple zeroes.

This time the appropriate ``harmonic'' form is $\Omega_2$ because
we have :
\begin{eqnarray}
h_1(\theta) &\sim&  -\coeff{1}{2} \, \alpha_1  \,, \quad
h_2(\theta) ~\sim~   \coeff{1}{4} \, \alpha_2 \, \hat r  \,
e^{-{\hat r}} \,, \quad h_ 3 (\theta) ~\sim~  \coeff{1}{4} \,
\alpha_3 \, \hat r  \, e^{-{\hat r}}  \,, \qquad \theta \to 0 \,;
\nonumber\\
h_1(\theta) &\sim&  - \coeff{1}{4} \, \alpha_1 \,  r  \, e^{-{ r}}
\,, \quad h_2(\theta)  ~\sim~  - \coeff{1}{4} \, \alpha_2 \,  r
\, e^{-{  r}} \,, \quad h_ 3 (\theta) ~\sim~  \coeff{1}{2} \,
\alpha_3 \,, \qquad \theta \to \pi \,; \label{APhjasymp:1}
\end{eqnarray}
and so $\Omega_2$ is the only solution that falls off in both
asymptotic regions. It is, however, not really harmonic in that it
is singular precisely on the critical surface where $w_2 =0$.
This is, however, the standard behavior for the flux that goes
into making the complete, five-dimensional solution and, as was
noted in (\ref{AIform}), the complete flux, $C^{(3)}$, is smooth
on the critical surface.

One now has
\begin{equation}
Z    ~=~  \delta ~+~ \gamma \, \eta ~-~   {\alpha_2^2  \over w_2}
 \label{ZregAP}
\end{equation}
and $k = \mu \sigma_2$,  where
\begin{eqnarray}
\mu  &=&   -\coeff{3}{2}\, \delta\, h_2 ~-~ \coeff{3}{2}\,
\gamma\, \Big[ \, h_2 \eta  ~-~  \,  {\alpha_2^2 \over 4\, h_2 }
\,
\Big({1 \over w_2}  - \eta\Big)\Big]  \nonumber \\
&& ~-~ \, {\alpha_2^4 \over 8\, h_2}\,  \Big[ \,{2\, w_1 \, w_3
\over w_2^3}  ~-~ {w_1 +  w_3 \over w_2^2} \, \Big] ~+~ {\beta_2
\over   h_2 }   \,.
 \label{muegAP}
\end{eqnarray}

Recall that the vector potential for $\Omega_2$ is given in
(\ref{Bjdefn}) and so the potential for the complete Maxwell field
is:
\begin{equation}
A  ~=~   Z^{-1} \,\big(dt + \mu\,  \sigma_2 \big)   ~-~  h_2 \,
\sigma_2 \,. \label{ApotAP}
\end{equation}
and so the only potentially singular term is:
\begin{equation}
  Z^{-1} \,\mu     ~-~  h_2  ~\sim~ - {\alpha_2 \over 4\, u^2\, w_2} \,
  (4\, w_1 \, w_3   ~+~    u^4   )\,,
\label{potsing}
\end{equation}
as $w_2 \to 0$. However, from (\ref{DHsol}) one has
\begin{equation}
 w_1 \, w_3   ~+~   \coeff{1}{4}\,  u^4 ~=~  w_2  \, (w_1+ w_3) - w_2^2\,,
\label{wident}
\end{equation}
 and so the complete Maxwell field is regular.

The spatial sections of the complete five-dimensional metric are:
\begin{equation}
ds^2 ~=~  \big(\coeff{1}{4} \, b^2\, Z ~-~  \mu^2\, Z^{-2} \big)\,
\sigma_2^2 ~+~
 \coeff{1}{4} \, Z \,a^2\, b^2\, c^2 \, d\eta^2 ~+~ \coeff{1}{4} \,Z\, a^2\sigma_1^2 ~+~
 \coeff{1}{4} \, Z\, c^2\sigma_3^2 \,.
\label{APssmetric}
\end{equation}
First note that:
\begin{eqnarray}
 Z\, a^2 &=&  {w_3\over w_1}\,((\delta ~+~ \gamma \, \eta)\,w_2  ~-~    \alpha_2^2)  \,, \qquad
 Z\, c^2 ~=~  {w_1\over w_3}\,((\delta ~+~ \gamma \, \eta)\,w_2  ~-~    \alpha_2^2)  \,, \nonumber \\
 Z\, a^2\, b^2 \, c^2 &=&  \, w_1 \,w_3( (\delta ~+~ \gamma \, \eta) \, w_2  ~-~  \alpha_2^2)   \,.
\label{Zmetcoeff}
\end{eqnarray}
Since one has $w_1 <0$ and $w_3 >0$ everywhere (see
Fig.\ref{wjplots}) it follows that these three metric coefficients
are regular and positive near $w_2 =0$.

More generally, observe that $\delta + \gamma \eta \to \delta $
and $w_2 \to -\infty$ as  $\theta \to \pi$ and  $\delta + \gamma
\eta \to \delta -2\pi \gamma $ and $w_2 \to + \infty$ as $\theta
\to 0$.  This means that for the metric coefficients in
(\ref{Zmetcoeff}) to remain positive at infinity one must have:
\begin{equation}
\gamma ~\ge~   {\delta  \over 2 \pi}  ~\ge~ 0     \,.
\label{gamcond}
\end{equation}
Indeed observe that the function, $ \eta + \coeff{1}{2} \eta_0$,
is odd under $\theta \to \pi - \theta$ and so, for $\gamma >0$,
the function
\begin{equation}
 \gamma( \eta + \coeff{1}{2} \eta_0 )\, w_2 ~=~ \gamma( \eta + \pi)\, w_2
\label{nicefunction}
\end{equation}
is globally negative with a double zero at $\theta = {\pi \over
2}$.  Thus the metric coefficients (\ref{Zmetcoeff}) are globally
positive when $\delta$ is
 the middle  of the range specified by (\ref{gamcond}).

Now consider the remaining coefficient, $Z^{-2} \cQ$, where
\begin{equation}
\cQ~\equiv~   \coeff{1}{4} \, b^2\, Z^3 ~-~  \mu^2 \,.
\label{CTCprobAP}
\end{equation}
Near $w_2 =0$ one has $Z^{-2} \sim \alpha_2^{-4} w_2^2$ and
\begin{equation}
 \cQ ~\sim~  {\alpha_2^6 \,w_1 \,w_3 \over u^4\, w_2^4}\, \big(   w_2 \, (w_1+ w_3)  ~-~
 (w_1 \, w_3 +  \coeff{1}{4}\, u^4) \big) ~+~ \cO(w_2^{-2}) \,.
\label{cQasympAP}
\end{equation}
However, it follows from (\ref{wident}) that, in fact,  $\cQ  \sim
\cO(w_2^{-2})$ and so the metric coefficient  $Z^{-2} \cQ$ is
regular around $w_2 =0$.

The regularity of the solution near the critical surface was, of
course, guaranteed by our general analysis of the Toda metrics in
Section \ref{GenGeoms}, but it is still useful to see how it comes
about here.

Finally there is the angular momentum vector and the issue of
global positivity of $\cQ$.  For this it is most convenient to
consider the combination $h_2 \, \mu$:
\begin{eqnarray}
h_2\, \mu  &\sim&   \coeff{3}{8} \, \gamma\,\eta_0 \, \alpha_2^2
~+~ \beta_2 ~+~
\coeff{1}{8} \, \pi^2\, \alpha_2^4~+~ \cO( \theta^2) \,,  \qquad \theta \to 0 \,,\\
h_2\, \mu  &\sim&  \beta_2 ~-~ \coeff{1}{8} \, \pi^2\,
\alpha_2^4~+~ \cO((\pi-\theta)^2) \,, \qquad \theta \to \pi \,.
 \label{htwomu}
\end{eqnarray}
Since $h_2$ vanishes exponentially fast  in $r$ and $\hat r$  in
the two asymptotic regions (see (\ref{APhjasymp:1})), this means
that $\mu$ will diverge exponentially in $r$ and $\hat r$ unless
\begin{equation}
\beta_2~=~ \coeff{1}{8} \,  \pi^ 2\,\alpha_2^4  \,, \qquad
\gamma~=~ -\coeff{2}{3} \, \pi^ 2\,\eta_0^{-1}\, \alpha_2^2~=~
-\coeff{1}{3} \,   \pi \,  \alpha_2^2  \,.
 \label{betagammafix}
\end{equation}
If these two conditions are met then $\mu$ also vanishes
exponentially in $r$ and $\hat r$ in both of the asymptotic
regions.

Unfortunately this value of $\gamma$ is inconsistent with
(\ref{gamcond}).  If one allows $\mu$ to diverge exponentially in
one of the asymptotic regions then $\cQ$ will become negative in
the asymptotic regions.  This is because $Z$ limits to a finite
value and $b^2$ diverges as a power of $r$ or $\hat r$.  Therefore
there is no way to arrange the metric to be positive definite in
the asymptotic regions on both sides of the wormhole: Either one
has (\ref{gamcond}) and arranges that three coefficients in
(\ref{Zmetcoeff}) to be globally positive, or one arranges that
$\cQ >0$ only to have the three coefficients in (\ref{Zmetcoeff})
to change sign in one of the asymptotic regions.

Thus we have a beautifully regular metric across the critical
surface, but it fails to be globally well-behaved as a
``wormhole'' metric.  We suspect that the problem is due to the
high level of symmetry.  With more bubbles and thus more
parameters we believe that one could simultaneously control
behavior in both asymptotic regions.  Even with the very high
level of symmetry, there is another way to remove the regions of
CTC's.

\subsection{Pinching off the wormhole}

One way to remove the region of CTC's is to pinch off the wormhole
before one encounters the region where CTC's occur.  Here we will
consider the ambi-polar metric described exactly as above with the
asymptotic regions as $\theta \to \pi$ arranged to be regular and
asymptotic to the $U(1)$ fibration over $\IR^3$ as in
(\ref{APasympmet}).  This requires one to take:
\begin{equation}
\beta_2~=~ \coeff{1}{8} \,  \pi^ 2\,\alpha_2^4  \,, \qquad
\delta~>~ 0   \,.
 \label{betapinch}
\end{equation}

The metric coefficients, $a_i^2$, are non-vanishing away from the
critical surface, and so to pinch off the complete metric away
from the critical surface we must arrange that  the function $Z$
vanish at some point. To avoid CTC's one must also ensure that
$\cQ$ is non-negative near the pinch-off and so one must arrange
that  $\mu$ vanishes simultaneously with $Z$.   Thus we are
looking for a point, $\theta_0$, such that
\begin{equation}
Z\big|_{\theta = \theta_0}~=~ 0  \,, \qquad \mu |_{\theta =
\theta_0}~=~ 0   \,.
 \label{pinchconds}
\end{equation}
Given these conditions, the equation of motion, (\ref{BPSAH:3}),
for $\mu$ then implies that $\frac{d}{d \theta} \mu$ must also
vanish at $\theta_0$. Therefore, near the pinching-off point we
have:
\begin{equation}
Z ~\sim~ z_0 \, (\theta-  \theta_0) \,, \qquad  \mu ~\sim~ \mu_0
\, (\theta-  \theta_0)^2 \,, \qquad  \big(\coeff{1}{4} \, b^2\, Z
~-~ \mu^2\, Z^{-2} \big) ~\sim~ \coeff{1}{4} \, b_0^2 \, z_0
(\theta-  \theta_0)   \,.
 \label{pinchasymp}
\end{equation}
This means that the spatial part of the complete metric
(\ref{APssmetric}) is indeed pinching off in every direction with
surfaces of constant $\theta$ being a set of collapsing, squashed
three-spheres.  The metric is not smooth at $\theta_0$: There is a
curvature singularity in the spatial part of the metric and the
coefficient of $dt^2$ is diverging as $ (\theta- \theta_0)^{-2}$.
This reflects a similar divergence in the electric component of
the Maxwell fields, $A^{(I)}$, (see (\ref{Thetadefn})) at $Z = 0$.
One should also note that the flux, $\Theta$, is also singular at
$Z = 0$ in that it remains constant while the cycle that supports
it is collapsing.

Define
\begin{equation}
\hat \gamma ~\equiv~  \alpha_2^{-2}\,  \gamma \,, \qquad \hat
\delta ~\equiv~ \alpha_2^{-2} \delta   \,,
 \label{hatparams}
\end{equation}
then the conditions (\ref{pinchconds}) relate $\hat \gamma$ and
$\hat \delta$ to $\theta_0$.  Thus we can, in principle, choose
the pinching-off point and then (\ref{pinchconds}) yields the
corresponding values of $\hat \gamma$ and $\hat \delta$.  In
practice, there is the constraint that $\delta > 0$. We know from
the analysis above that we cannot arrange for $Z$ and $\mu$ to
vanish simultaneously at $\theta =0$.   Numerical analysis shows
that one cannot  have  $Z$ and $\mu$ vanish simultaneously unless
$\theta_0 \gtrsim 0.6158$.    Since we are interested in solutions
that contain the critical surface ($w_2 =0$), we have found a
number of solutions that pinch off  for $ 0.6158 \lesssim \theta_0
< \frac{\pi}{2}$. We also checked numerically that it does not
appear to be possible to have all three of $\mu$, $Z$ and ${d Z
\over d \theta}$ vanish simultaneously for $\theta \in (0,
\frac{\pi}{2})$.  Thus (\ref{pinchasymp}) appears to be the
general behavior at a pinch-off: $Z$ does not appear to be able to
have a double root.

\begin{figure}
\centering
\centerline{ {\includegraphics[height=5cm]{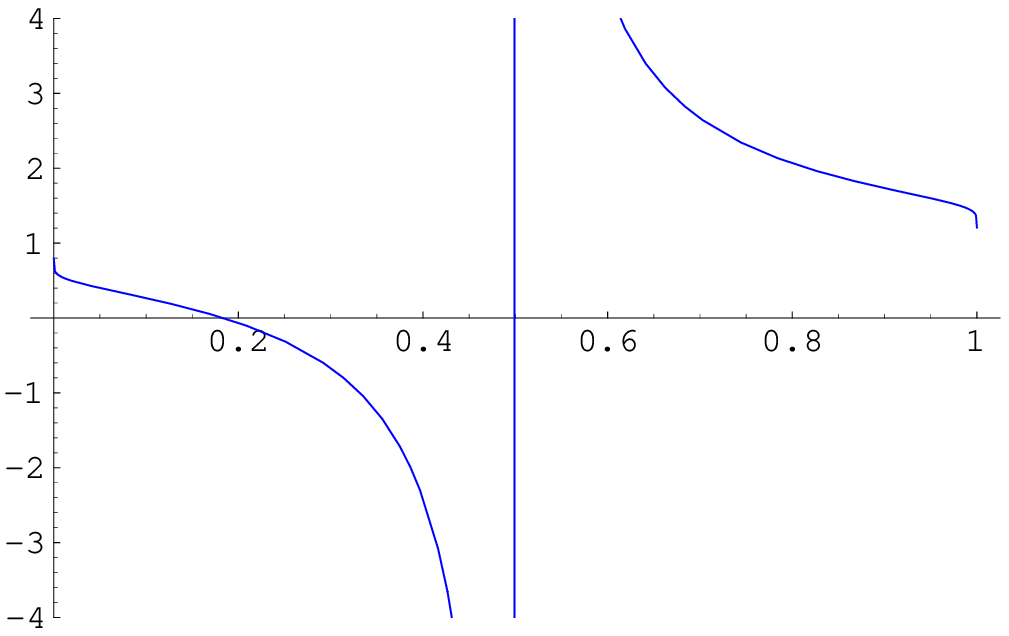} \hskip 1cm
 \includegraphics[height=5cm]{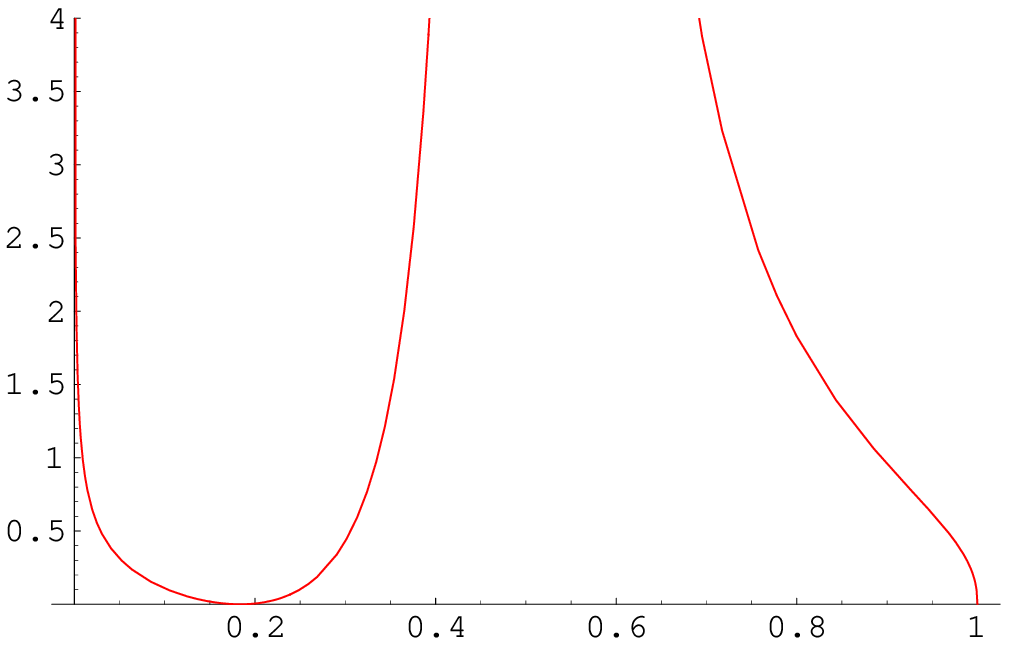}}}
\caption{ These are graphs of $Z$ (on the left) and $\mu$  (on the
right)  as  functions of $x = \sin^2\coeff{\theta}{2}$ for
$\gamma =0$, $\delta =1$ and $\alpha_2 \approx 0.4890$.  Both
functions are singular at $x=0.5$ and both vanish, $\mu$ with a
double root, at $x  \approx 0.1837$.}
\label{ZMupics}
\end{figure}

\begin{figure}
\centering
\includegraphics[height=6cm]{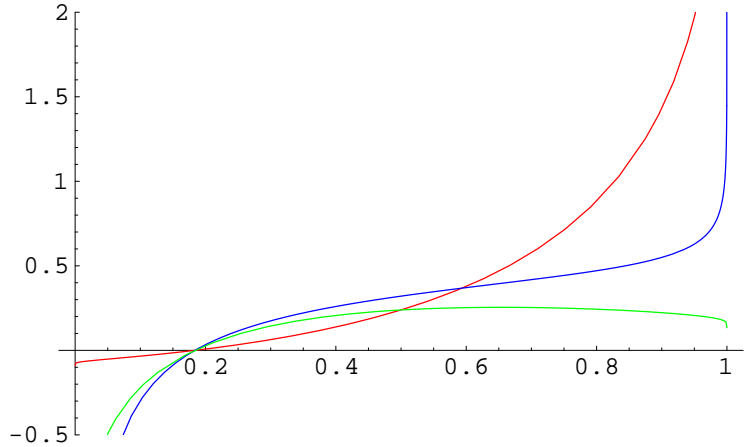}
\caption{This graph shows the three metric coefficients in the
angular directions, $Z a^2$, $Z^{-2} \cQ$ and $Z c^2$ (in this
order from top to bottom on the right-hand side of the graph), as
functions of $x = \sin^2\coeff{\theta}{2}$ for  $\gamma =0$,
$\delta =1$ and  $\alpha_2 \approx 0.4890$. All of these functions
vanish at the pinching-off point, $x  \approx 0.1837$, and are
positive-definite to the right of it.}
\label{metpic}
\end{figure}

We have verified in several numerical examples that the spatial
metric is indeed globally positive definite in the region at and
to the right of the pinch. These solutions still contain the
critical surface where the $a_i^2$ and $Z$ simultaneously change
sign and these solutions are perfectly regular across the critical
surface.  The cost of ensuring the global absence of CTC's is to
include a non-standard, singular point-source at the center of the
solution.

To present an example, we considered the solution with $\gamma
=0$, $\delta =1$. Solving (\ref{pinchconds}) leads to $\alpha_2
\approx 0.4890$ and the pinch-off at  $x =
\sin^2(\frac{\theta}{2}) \approx 0.1837$.  In Fig.  \ref{ZMupics}
we show plots of the functions $Z$ and $\mu$ for these parameter
values. Note that both are singular at $x=0.5$ and that both
vanish, $\mu$ with a double root, at $x  \approx 0.1837$.  In Fig.
\ref{metpic} we have shown the three metric coefficients in the
angular directions, $Z a^2$, $Z c^2$ and $Z^{-2} \cQ$.  All of
them are positive and vanish exactly at the pinching-off point.

Before ending this section we should make a few more comments
about the metric that is pinching off. The singularity at the
pinch-off point is caused by the fact that the warp factors $Z_I$
go to zero. This causes the size of the two-cycles wrapped by
fluxes to shrink to zero size, and hence the energy density coming
from these fluxes to be infinite. A well-known solution with a
similar type of singularity is the one obtained by Klebanov and
Tseytlin \cite{Klebanov:2000nc}. However,  for this solution it is
well understood that the singularity comes about because of the
high level of symmetry in the Ansatz, and that upon considering a
less-symmetric base space the singularity is resolved
\cite{Klebanov:2000hb}. Since the base space considered here also
has a high level of symmetry, it is tempting to conjecture that,
in analogy to the Klebanov-Strassler solution
\cite{Klebanov:2000hb}, the pinching off will be resolved by the
blowing up of a two-cycle on the base, which will only be possible
in a less-symmetric, non-singular background.

We should also remark that in our discussion we have taken all
three warp factors to be equal, but generically we can also
imagine pinching off the metric using only one of the warp
factors, and keeping the others finite. This will change the
structure of the metric near the singularity (some of the two-tori
will blow up and some others will shrink), but the singularity
will also come from shrinking cycles on the base, and will
probably be resolved also by considering a less-symmetric base
with a blown-up two-cycle

\section{Variations on the Eguchi-Hanson metric}

Given the foregoing results, particularly those involving
wormholes, it is interesting to look at the corresponding story
for the Eguchi-Hanson metric \cite{Eguchi:1979yx}.  This metric
has an $SO(3) \times U(1)$ invariance and the diagonal $U(1)$
action is triholomorphic.  The metric is equivalent to a GH metric
with two GH points of  equal charge \cite{Prasad:1979kg}.  The
manifestly $SO(3) \times U(1)$ invariant form of this metric is:
\begin{equation}
ds^2 = \bigg(1- \frac{a^4}{\rho^4} \bigg)^{-1} \, d\rho^2  ~+~
\frac{\rho^2}{4} \bigg(1- \frac{a^4}{\rho^4} \bigg)\sigma_3^2~+~
\frac{\rho^2}{4} (\sigma_1^2 + \sigma_2^2) \,. \label{EHstand}
\end{equation}
The space contains an $S^2$ (bolt) at $\rho = a$ and so the range
of the radial coordinate is $a \le \rho < \infty$.  At infinity
this space is asymptotic to $\IR^4/\ZZ_2$.

To avoid closing off of the space at the bolt, we analytically
continue by taking $ a^2 = i b$, with $b$ real, and   introduce a
new radial coordinate $\eta = \rho^2$. One thereby obtains:
\begin{equation}
ds^2 = \bigg(1+ \frac{b^2}{\eta^2}
\bigg)^{-1}\frac{d\eta^2}{4\eta} +~ \frac{\eta}{4}(\sigma_1^2 +
\sigma_2^2) +~ \frac{\eta}{4} \bigg(1+  \frac{b^2}{\eta^2} \bigg)
\sigma_3^2\,. \label{newEH}
\end{equation}

This metric was also  considered by Eguchi and Hanson in
\cite{Eguchi:1978xp}, where it was called ``type I,''  and was
given in the form:
\begin{equation}
ds^2 = \left( 1 +
\left(1-\ds\frac{a^4}{r^4}\right)^{-1/2}\right)^2
\ds\frac{dr^2}{4} + \ds\frac{r^2}{8}\left( 1 +
\left(1-\ds\frac{a^4}{r^4}\right)^{1/2}\right)(\sigma_1^2 +
\sigma_2^2) + \ds\frac{r^2}{4}\sigma_3^2 \,. \label{EHtypeI}
\end{equation}
This may be mapped to (\ref{newEH}) via the coordinate change
\begin{equation}
\eta ~=~   r^2\left( 1 + \sqrt{1-\ds\frac{a^4}{r^4}}\right)  \,.
\end{equation}

In terms of the Toda frame, (\ref{Todametric})--(\ref{TodaEqn}),
this metric was found in \cite{DasGegenberg} and is given by
\begin{equation}
\nu = \log\left(z^2+\frac{a^4}{16}\right)- \log(2)-
2\log\left(1+\frac{x^2+y^2}{8}\right) \,.
\end{equation}
\\
The reason why this metric was never studied in the past is that
it is not geodesically complete, and there is a singularity at
$\eta=0$. Nevertheless, we can extend the coordinate $\eta =
\rho^2$ to negative values, and the resulting space (\ref{newEH})
has two regions, one where the signature is $(+,+,+,+)$ and one
where the signature is $(-,-,-,-)$ . This makes (\ref{newEH}) into
precisely an ambi-polar metric of the type that can give a good
five-dimensional BPS solution: The overall sign of the metric
changes as one passes through $\eta =0$, with the coefficient of a
$U(1)$ fiber becoming singular at this critical surface.

\subsection{The BPS solutions}

In order to solve the BPS equations,
(\ref{BPSeqn:1})--(\ref{BPSeqn:3}), it is convenient to introduce
the basis of frames given by:
\begin{equation}
\hat{e}_1=-\left(1+\frac{b^2}{\eta^2}\right)^{-1/2}\frac{d\eta}{2\sqrt{\eta}}\,,
\qquad \hat{e}_2=\frac{\sqrt{\eta}}{2}\sigma_1 \,, \qquad
\hat{e}_3=\frac{\sqrt{\eta}}{2}\sigma_2\,, \qquad
\hat{e}_4=\ds\frac{\sqrt{\eta}}{2}\left(1+\frac{b^2}{\eta^2}\right)^{1/2}\sigma_3
\end{equation}
One can then show that:
\begin{equation}
\Theta=\frac{\alpha}{\eta^2}(\hat{e}_1\wedge\hat{e}_4+\hat{e}_2\wedge\hat{e}_3)
\end{equation}
defines a harmonic, self-dual, ``normalizable'' two form for
constant $\alpha$.  One also has $\Theta=dB$ with
$B=\ds\frac{\alpha}{4\eta}\sigma_3$.  As before, we take all three
flux forms $\Theta^{(I)}$ to be equal to $\Theta$ and set $Z_I
=Z$. Then the equation for $Z(\eta)$ becomes
\begin{equation}
\frac{d}{d\eta}\left((\eta^2+b^2)\frac{dZ}{d\eta}\right)=\frac{\alpha^2}{2\eta^3}
\end{equation}
which is solved by
\begin{equation}
Z(\eta)=\gamma +
\frac{\alpha^2}{4b^2\eta}+\left(\frac{\alpha^2}{4b^3}+\frac{\beta}{b}\right)
\arctan\left(\frac{\eta}{b}\right) \,, \label{ZEHform}
\end{equation}
where $\beta$ and $\gamma$ are integration constants. The angular
momentum vector, $k$, has a solution of the form
$k=\mu(\eta)\sigma_3$ where the function $\mu(\eta)$ satisfies
\begin{equation}
\eta^3\frac{d\mu}{d\eta}-\eta^2\mu
+\frac{3\alpha\gamma}{4}\eta+\frac{3\alpha}{4}\left(\frac{\alpha^2}{4b^3}+
\frac{\beta}{b}\right)\arctan\left(\frac{\eta}{b}\right)+\frac{3\alpha^3}{16b^2}=0\,.
\end{equation}
The solution to this equation is
\begin{multline}
\mu(\eta) = \delta\eta
+\frac{3\alpha}{8b^2}\left(\frac{\alpha^2}{4b^2}+\beta\right) +
\frac{3\alpha\gamma}{8}\frac{1}{\eta} +
\frac{\alpha^3}{16b^2}\frac{1}{\eta^2} +
\frac{3\alpha}{8b^3}\left(\frac{\alpha^2}{4b^2}+
\beta\right)\eta\arctan\left(\frac{\eta}{b}\right)
+\\
\frac{3\alpha}{8b}\left(\frac{\alpha^2}{4b^2}+\beta\right)\frac{1}{\eta}
\arctan\left(\frac{\eta}{b}\right) \,. \label{muEHform}
\end{multline}
To complete the solution we have to impose boundary condition on
the functions $Z$ and $\mu$.

\subsection{A regular ``wormhole''}

If the solution is to have two asymptotic regions corresponding to
$\eta\rightarrow\pm\infty$ then we must require that the angular
momentum vector falls off in these regions or there will
generically be CTC's. This implies:
\begin{equation}
\delta = 0 \,, \qquad\qquad \beta=-\ds\frac{\alpha^2}{4b^2} \,,
\end{equation}
and then the functions  $Z$ and $\mu$ simplify to:
\begin{equation}
Z(\eta)=\gamma + \frac{\alpha^2}{4b^2\eta} \,, \qquad \mu(\eta) =
\frac{3\alpha\gamma}{8}\frac{1}{\eta} +
\frac{\alpha^3}{16b^2}\frac{1}{\eta^2} \,. \label{ZmuEHform}
\end{equation}
If $\gamma\ne 0$, $Z$ will have a zero at  $\eta =
-\frac{\alpha^2}{4b^2\gamma}$ and thus we will inevitably have
CTC's unless we pinch off the solution before, or at, this point.

We consider $\gamma=0$ first, for which we have:
\begin{equation}
Z = \ds\frac{\alpha^2}{4b^2}\frac{1}{\eta} \,, \qquad\qquad \mu =
\ds\frac{\alpha^3}{16 b^2} \frac{1}{\eta^2} \,. \label{ZandmuwhEH}
\end{equation}
Note that the angular momentum function $\mu$ is always positive
and is diverging on the critical surface $\eta=0$.  The $Z$ also
diverges and changes sign on the critical surface. This behavior
ensures that the five-dimensional metric  is regular and
Lorentzian. The explicit form of the space-time metric is:
\begin{equation}
ds^2_5 = -\ds\frac{16b^4}{\alpha^4}\eta^2dt^2 -
\ds\frac{2b^2}{\alpha}dt\,\sigma_3 +
\ds\frac{\alpha^2}{16b^2}\ds\frac{d\eta^2}{(\eta^2+b^2)} +
\ds\frac{\alpha^2}{16b^2}(\sigma_1^2+\sigma_2^2+\sigma_3^2)\,.
\label{womrholeEH}
\end{equation}
This metric can be cast into a more familiar form by first
diagonalizing the metric by shifting the $\psi$-coordinate in
(\ref{sigmadefn}) so that $\sigma_3\rightarrow\sigma_3+
\ds\frac{16b^4}{\alpha^3}dt$:
\begin{equation}
ds^2_5=-\ds\frac{16b^4}{\alpha^4}(\eta^2+b^2)dt^2+\ds\frac{\alpha^2}{16b^2}
\ds\frac{d\eta^2}{\eta^2+b^2}+\ds\frac{\alpha^2}{16b^2}(\sigma_1^2+
\sigma_2^2+\sigma_3^2)\,.
\end{equation}
Change variables via $\eta  =b \sinh\chi$,
$\tilde{t}=\ds\frac{16b^4}{\alpha^3}t$ and then the metric becomes
\begin{equation}
ds^2_5=\ds\frac{\alpha^2}{16b^2}\, (-\cosh^2\chi \,
d\tilde{t}^2+d\chi^2+\sigma_1^2+\sigma_2^2+\sigma_3^2)\,,
\end{equation}
which is the well known metric  for global $AdS_2\times S^3$. The
complete Maxwell field on this space is given by
\begin{equation}
 dA  ~=~ d\Theta -  d(Z^{-1}(dt+k)) \,,
\end{equation}
and, using $\Theta=dB$ with $B = \ds\frac{\alpha}{4\eta}\sigma_3$,
we find
\begin{equation}
A ~=~  - \ds\frac{4b^2}{\alpha^2}\, \eta\, dt \,, \qquad F ~=~
\ds\frac{4b^2}{\alpha^2}\, dt \wedge d\eta \,.
\end{equation}
The Maxwell field is thus proportional to the volume form on
$AdS_2$ and we have obtained the global form of a
Robinson-Bertotti solution. The wormhole  thus reduces to the
usual global AdS solution.

In fact the solution based on the singular EH metric (\ref{newEH})
(with $\eta>0$) has also been discussed in \cite{5dsols}, where it
was shown to give an $AdS_2 \times S^3$ solution with the $AdS_2$
in global coordinates. What we find very puzzling is the fact that
in order to get the {\it entire} range of coordinates for the
global $AdS_2$ metric, one must start from the ambi-polar EH
metric with the coordinate $\eta$ running between $-\infty$ and
$\infty$.

It is interesting to try to understand the reason for which we
could find an Eguchi-Hanson ``wormhole'' but not an Atiyah-Hitchin
one. At an
 algebraic level, the problem comes from the form of $\mu$, which
in the Eguchi-Hanson background goes to zero on both asymptotic
regions (\ref{ZmuEHform}), while in the Atiyah-Hitchin background
$\mu$,  (\ref{muegAH}), diverges in one region or in the other. If
one relaxes the requirement that the Atiyah-Hitchin solutions be
asymptotically flat, one can choose a more generic $\Theta$,
containing all three $\Omega_i$. However, this still does not give
a $\mu$ that decays properly at the two asymptotic regions.

\subsection{A ``pinch-off'' solution}

The other way to remove CTC's is to allow $\gamma \ne 0$ in
(\ref{ZEHform}) and pinch-off  the asymptotic region with $\eta
\to - \infty$ at the point, $\eta_0$, where $Z$ vanishes. This
means that we only have to require that $\mu$ vanish  as
$\eta\rightarrow\infty$ and  this implies
\begin{equation}
\delta =
-\frac{3\pi\alpha}{16b^3}\left(\beta+\frac{\alpha^2}{4b^2}\right)
\end{equation}
in (\ref{muEHform}).

As with the Atiyah-Hitchin solution, the solution will have CTC's
near the pinching off point unless we also require that $\mu$
vanishes at the same point.  Specifically, the constant time
slices of the metric have the form:
\begin{equation}
ds^2 = -\frac{\mu^2}{Z^2}\,\sigma_3^2 +
Z\left(\left(1+\frac{b^2}{\eta^2}\right)^{-1}\frac{d\eta^2}{4\eta}
+ \frac{\eta}{4}(\sigma_1^2 + \sigma_2^2) +
\frac{\eta}{4}\left(1+\frac{b^2}{\eta^2}\right) \sigma_3^2\right)
\end{equation}
and to avoid  CTC's we must have
\begin{equation}
\cQ ~\equiv~ Z^3\left(\frac{\eta^2+b^2}{4\eta}\right)-\mu^2 ~\ge~
0 \,.
\end{equation}
If $Z$ vanishes then $\mu$ must vanish and this imposes a
relationship, akin to the bubble equations, on $\beta$, $\alpha$
and $b$.  Unlike the corresponding solution in the Atiyah-Hitchin
background, there is still a free parameter in the final result,
and if one choses these parameters in the proper ranges one can
arrange that the pinch-off occurs at $\eta_0 < 0$ and that there
are no CTC's in the region $\eta > \eta_0$.  There is still,
however, a curvature singularity in the metric at $\eta = \eta_0$,
similar to the one in the pinched-off Atiyah-Hitchin solution, and
probably caused also by the fact that the ansatz used is very
symmetric. It is quite likely that this singularity will also be
resolved in the same manner as the
Klebanov-Tseylin/Klebanov-Strassler solutions
\cite{Klebanov:2000nc,Klebanov:2000hb}

It is easy to find numerical examples that exhibit a ``pinch
off.'' For example, one can take the following values of the
parameters:
\begin{equation}
\alpha\approx4.2619 \qquad \beta\approx-4.5358 \qquad \gamma=b=1
\qquad \text{and} \qquad\delta =
-\frac{3\pi\alpha}{16b^3}\left(\beta+\frac{\alpha^2}{4b^2}\right)
\,,
\end{equation}
and the pinch off point   is $\eta_0\approx-4.5721$.
 Since $\eta_0$ is negative this represents a solution based upon a
 non-trivial  ambi-polar base metric.

\section{Conclusions}

We have investigated the construction of three-charge solutions
that do not have a tri-holomorphic $U(1)$ isometry.  We have found
that the most general form of these solutions, can be expressed in
term of several scalar functions. One of these functions satisfies
the (non-linear) $SU(\infty)$ Toda equation, while the other
functions satisfy linear equations that can be thought of as
various linearizations of the $SU(\infty)$ Toda equation.

We have also shown generically that in the region where the
signature of the four-dimensional base space changes from
$(+,+,+,+)$ to $(-,-,-,-,)$, the fluxes, warp factors, and the
rotation vector diverge as well, but the overall five-dimensional
(or eleven-dimensional) solution is smooth. This is similar to
what happens when the base-space is Gibbons-Hawking, and strongly
suggests that this phenomenon is generic: {\it Any
ambi-polar\footnote{As explained in the bulk of this
    paper, an ambi-polar metric is one whose signature changes from
    $(+,+,+,+)$ to $(-,-,-,-,)$, such that the three-metric on the
    critical surface has two vanishing eigenvalues and one divergent
    eigenvalue.}, four-dimensional, hyper-K\"ahler metric with at
  least one non-trivial two-cycle can be used to construct a regular
  supersymmetric five-dimensional three-charge solution upon adding
  fluxes, warp factors and rotation according to the BPS equations
  (\ref{BPSeqn:1}), (\ref{BPSeqn:2}) and (\ref{BPSeqn:3}).}

This phenomenon is likely to be quite important in the programme
of establishing whether black holes are ensembles of smooth
supergravity or string solutions. To prove this conjecture, one
would need to construct and count smooth horizonless solutions
that have the same charges and angular momenta as three-charge
black holes or black rings. Our analysis suggests that this
counting problem is in fact much easier, since one would not have
to count the full solutions, but just the hyper-K\"ahler base
spaces underlying them.  It also suggests that we may be able to
capture the essential structure of the hyper-K\"ahler base by
approximating it with a quilt of GH spaces.

Since the most general form of hyper-K\"ahler, four-dimensional
spaces with a rotational $U(1)$ isometry is not known explicitly,
one cannot explicitly construct the most general three-charge
bubbling solution with this isometry. Nevertheless, we have been
able to construct a first explicit bubbling solution with a
rotational $U(1)$ starting from an ambi-polar generalization of
the Atiyah-Hitchin metric. For both the standard Atiyah-Hitchin
and Eguchi-Hanson metrics, it is not possible to construct regular
three-charge bubbling solutions. This reflects the fact that
fluxes tend to stabilize cycles that would shrink by themselves,
and hence only ``pathological'' generalizations to ambi-polar
metrics can be used as base-spaces to create bubbling solutions.
We have obtained the ambi-polar generalizations of both the
Atiyah-Hitchin and the Eguchi-Hanson spaces, and have constructed
the full three-charge solutions based on these spaces.

As expected from our general analysis, the full solutions are
completely regular at the critical surface where the metric on the
base space changes sign. Moreover, for the ambi-polar
Eguchi-Hanson space, one can construct the full solution, which,
interestingly enough, turns out to be {\it global} $AdS_2 \times
S^3$. We could also obtain solutions that pinch off, and have a
curvature singularity.  We argued that this singularity has the
same structure as the one in the Klebanov-Tseytlin solution
\cite{Klebanov:2000nc} and we believe the presence of this
singularity is a consequence of the high level of symmetry of the
base space, and that the singularity will similarly be resolved by
considering a less-symmetric base space.

This work opens several interesting directions of research. First,
having shown that singular, $U(1)$-invariant, ambi-polar,
four-dimensional, hyper-K\"ahler metrics can give smooth
five-dimensional solutions upon adding fluxes, it is important to
go back to the $SU(\infty)$ Toda equation and to construct more
general solutions. A first step in this investigation would be to
find the solutions of the Toda equations that give the $U(1)
\times U(1)$ invariant ambi-polar Gibbons-Hawking metrics,
following perhaps the techniques of \cite{Bakas:1996gf}. One could
then find other solutions in the vicinity of the latter, and count
them using the techniques of \cite{counting}.

Second, the fact that the ambi-polar generalizations of the
Atiyah-Hitchin and the Eguchi-Hanson spaces give regular
geometries suggests that ambi-polar generalizations of other known
hyper-K\"ahler metrics will also give regular solutions. Finding
these solutions would be quite interesting.

Finally, we have seen that the ambi-polar generalization of the
Eguchi-Hanson space yields a full geometry that is $AdS_2 \times
S^3$. Moreover, unlike in the case of usual bubbling BPS
solutions, the $AdS_2 $ solution is not the Poincar\'e patch, but
the full global $AdS$ solution.  While the distinction between
global and Poincar\'e $AdS_2$ is relatively trivial, the
appearance of something like a regular wormhole suggests that
bubbling geometries might be even richer and more interesting than
was originally anticipated.

\bigskip
\leftline{\bf Acknowledgements}
\smallskip
We would like to thank Juan Maldacena and Radu Roiban for
interesting discussions.  The work of NB and NW was supported in
part by funds provided by the DOE under grant DE-FG03-84ER-40168.
The work of IB was supported in part by the Dir\'ection des
Sciences de la Mati\`ere of the Commissariat \`a L'En\'ergie
Atomique of France. The work of NB was also supported in part by
the Dean Joan M. Schaefer Research Scholarship.



\end{document}